\pdfoutput=1
\documentclass[journal]{IEEEtran}
\usepackage{graphicx}
\usepackage{amsmath}
\usepackage{physics}
\usepackage{tabularx}

\usepackage{hyperref}

\usepackage{framed}
\usepackage{moreverb}
\usepackage[small,sf,bf]{subfigure}
\usepackage{xspace}
\usepackage{flushend}%

\usepackage{xfrac}%

\usepackage[ruled,vlined]{algorithm2e}

\usepackage{amsfonts}
\usepackage{amssymb}
\usepackage{amsmath}
\usepackage[thmmarks,amsmath,hyperref]{ntheorem}
\usepackage{graphicx}
\usepackage{times}
\usepackage[scaled]{helvet}%
\usepackage{textcomp}
\usepackage{color}
\usepackage{booktabs}
\usepackage{listings}
\usepackage{multicol}
\usepackage{times}
\usepackage{pifont}%
\usepackage{microtype}

\usepackage[scaled=0.85]{beramono}
\usepackage[T1]{fontenc}
\usepackage{url}
\usepackage{multirow}
\usepackage{array}

\newcommand\code[1]             {{\bf\ttfamily\small #1}}

\definecolor{listinggreen}{rgb}{0,0.6,0}
\definecolor{listinggray}{rgb}{0.5,0.5,0.5}
\definecolor{listingmauve}{rgb}{0.58,0,0.82}
\definecolor{listingkeywordcolor}{rgb}{1.0,0.4,0.0}
\definecolor{listinglightgray}{rgb}{0.9863,0.9863,0.9863}

\lstdefinelanguage{FSharp}
{
	morekeywords	= {
    let,
    type,
    Measure,
	},
	sensitive	= false,
	morecomment	= [l]{\#},
	morecomment	= [s]{(*}{*)},
}

\lstdefinelanguage{Newton}
{
	morekeywords	= {
		signal,
		derivation,
		symbol,
		name,
		invariant,
		constant,
		English,
		sensor,
		name,
		none,
		dot,
		cross,
		derivative,
		integral,
		interface,
		i2c,
		spi,
		analog,
		write,
		read,
		delay,
		range,
		erasuretoken,
		uncertainty,
		accuracy,
		precision,
		Gaussian,
		exponential,
		biexponential,
		to,
		bits,
		dimensionless,
		include
	},
	sensitive	= false,
	morecomment	= [l]{\#},
	morecomment	= [s]{/*}{*/},
}

\usepackage{hyperref}
\hypersetup{
    colorlinks=false, %
    linktoc=all,     %
    linkcolor=blue,  %
}

\usepackage{xcolor}
\usepackage{colortbl}

\definecolor{a}{rgb}{0.9,0.95,0.95}%
\definecolor{b}{rgb}{0.99,0.99,0.99}%
\definecolor{c}{rgb}{0.99,0.99,0.4}%

\newcommand\vectorSymbol[2]					{\mathbf{#1}_{#2}}

\newcommand\matrixSymbol[2]					{\mathbf{#1}_{#2}}
\newcommand\functionSymbol[1]				{\mathrm{#1}}

\newcommand\kalman							{{Kalman}\xspace}

\definecolor{mathHighlightPhysics}{rgb}{0.99,	0.75,	0.79}
\definecolor{mathHighlightNoise}{rgb}{	0.90,	0.90,	0.90}
\definecolor{mathHighlightPreSSA}{rgb}{	1.0,	1.00,	0.00}
\definecolor{mathHighlightPostSSA}{rgb}{0.08,	0.82,	1.00}

\newcommand{\highlightPhysics}[1]{\colorbox{mathHighlightPhysics}{$\displaystyle #1$}}
\newcommand{\highlightNoise}[1]{\colorbox{mathHighlightNoise}{$\displaystyle #1$}}
\newcommand{\highlightPreSSA}[1]{\colorbox{mathHighlightPreSSA}{$\displaystyle #1$}}
\newcommand{\highlightPostSSA}[1]{\colorbox{mathHighlightPostSSA}{$\displaystyle #1$}}

\newcolumntype{C}[1]{>{\centering\arraybackslash\hspace{0pt}}m{#1}}
\ifCLASSINFOpdf
\else
\fi

\begin{document}
\title{Automated Physics-Derived Code Generation for Sensor Fusion and State Estimation
}

\author{\IEEEauthorblockN{Orestis~Kaparounakis$^{\star}$\textsuperscript{\textsection}, Vasileios~Tsoutsouras$^{\dagger}$, Dimitrios~Soudris$^{\star}$, Phillip~Stanley-Marbell$^{\dagger}$}\\
\IEEEauthorblockA{$^{\star}$\textit{Microprocessors and Digital Systems Laboratory, National Technical University of Athens, Greece} \\
$^{\dagger}$\textit{Physical Computation Laboratory, University of Cambridge, United Kingdom} \\
$^{\star}$\{orestis.kapar, dsoudris\}@microlab.ntua.gr,
$^{\dagger}$\{vt298, phillip.stanley-marbell\}@eng.cam.ac.uk}%
\thanks{\textsuperscript{\textsection}This work started when O.~Kaparounakis was at the University of Cambridge.}%
\thanks{This research is supported by an Alan Turing Institute award TU/B/000096 under EPSRC grant EP/N510129/1, by EPSRC grant EP/R022534/1, and by EPSRC grant EP/V004654/1.}%
}%

\maketitle

\begin{abstract}
We present a new method for automatically generating the implementation
of state-estimation algorithms from a machine-readable specification
of the physics of a sensing system and physics of its signals and
signal constraints. We implement the new state-estimator code
generation method as a backend for a physics specification language
and we apply the backend to generate complete C code implementations
of state estimators for both linear systems (Kalman filters) and
non-linear systems (extended Kalman filters). The state estimator
code generation from physics specification is completely automated
and requires no manual intervention. The generated filters can
incorporate an Automatic Differentiation technique which combines
function evaluation and differentiation in a single process. 

Using the description of physical system of a range of complexities,
we generate extended \kalman filters, which we evaluate in terms of 
prediction accuracy using simulation traces.
The results show that our automatically-generated sensor fusion and
state estimation implementations provide state estimation within
the same error bound as the human-written hand-optimized counterparts. 
We additionally quantify the code size and dynamic instruction 
count requirements of the generated state estimator implementations on 
the RISC-V architecture. The results show that our synthesized
state estimation implementation employing Automatic Differentiation
leads to an average improvement in the dynamic instruction count
of the generated \kalman filter of 7\%--16\% compared to the standard
differentiation technique. This is improvement comes at the limited cost of 
an average 4.5\% increase in the code size of the generated filters.

\end{abstract}


\IEEEpeerreviewmaketitle

\section{Introduction}
\begin{figure}[t]
\centering
\subfigure[]{%
\includegraphics[width=0.50\columnwidth]{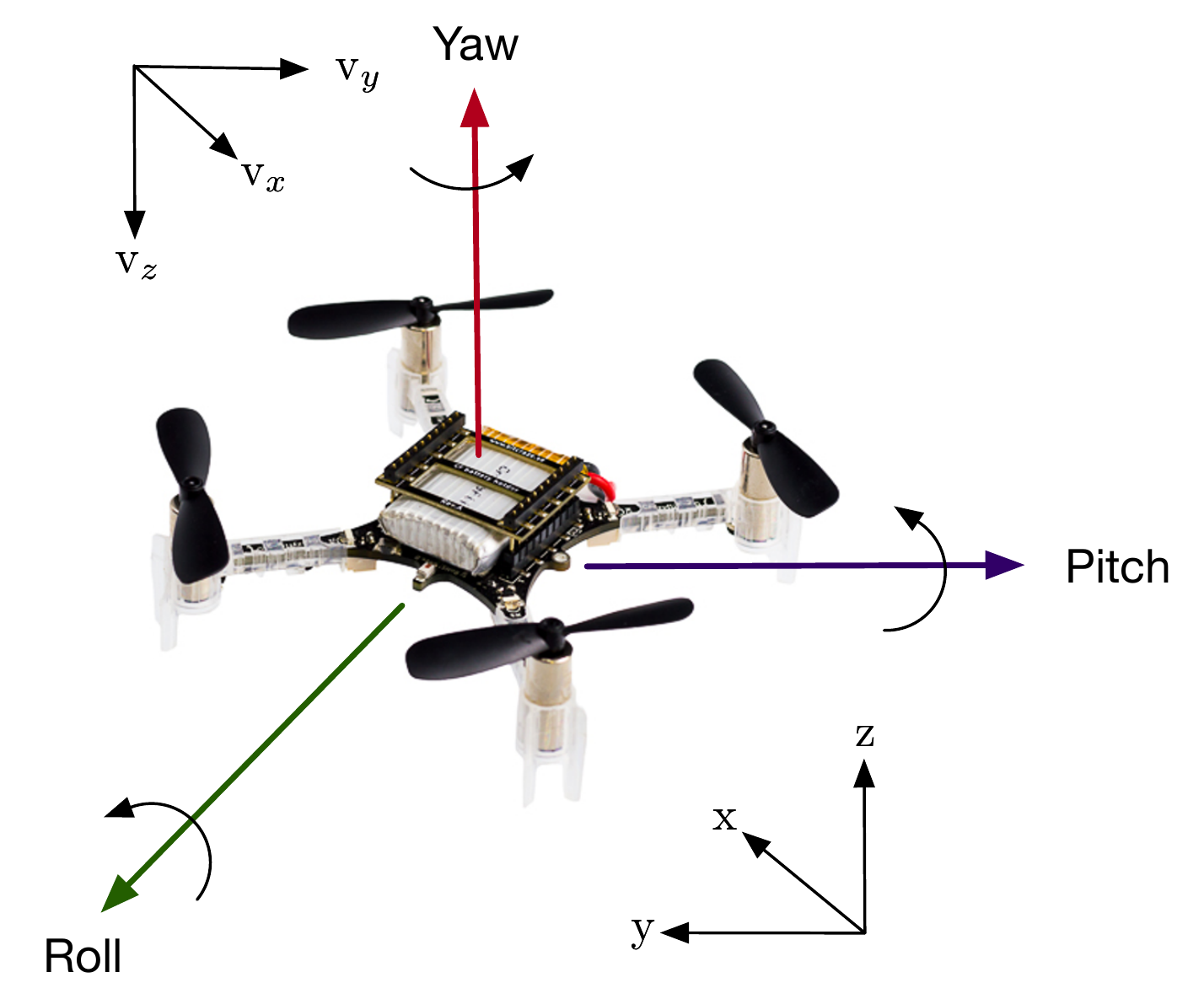}
\label{fig:subfigure1}}
\quad
\subfigure[]{%
\includegraphics[width=0.35\columnwidth]{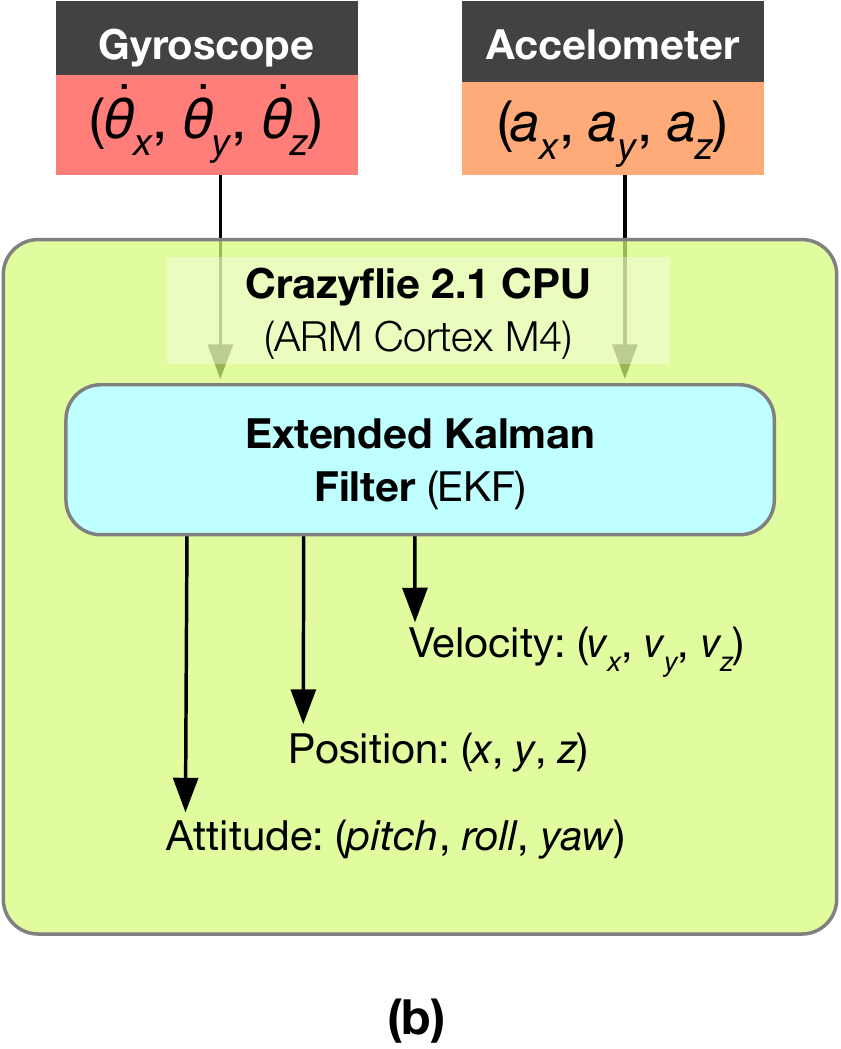}
\label{fig:subfigure2}}
\caption{\textbf{(a)} The CrazyFlie 2.1 micro-UAV. \textbf{(b)} The
micro-UAV combines noisy readings from its sensors (angular rate
$\dot{\theta}$ and acceleration $a$ in three dimensions) to obtain
a stable estimate of the state parameters $\mathit{pitch}$,
$\mathit{roll}$, $\mathit{yaw}$, position $(x, y, z)$, and velocity
$(v_x, v_y, v_z)$. }
\label{fig:introduction:applications-of-state-estimation}
\end{figure}

\IEEEPARstart{M}{any} computing systems use data from sensors to
drive control decisions. Because all measurements have some degree
of measurement uncertainty and sensors are no different, computing
systems that consume sensor data often use techniques ranging from
averaging or filtering~\cite{oppenheim1983digital}, to more
sophisticated state-estimation techniques~\cite{kalman19601960,
jazwinski1969adaptive, wan2000unscented, arulampalam2002tutorial},
to combine the signals from multiple sensors to obtain improved
noise rejection.
Figure~\ref{fig:introduction:applications-of-state-estimation} shows
one example system: A micro-unmanned aerial vehicle (micro-UAV)
weighing 27\,g whose control system comprises an ARM Cortex-M4
microcontroller with 192kB RAM~\cite{giernacki2017crazyflie}. The
system's flight control uses data from an inertial measurement unit
for 3-axis acceleration and 3-axis angular rate measurement as well
as a high precision pressure sensor for elevation monitoring. The
micro-UAV's control system uses a Kalman filter to fuse the
(noisy) readings from the seven dimensions of these sensors
to generate a stable pitch, roll, yaw, and elevation estimate in
real time~\cite{MuellerHamerUWB2015, MuellerCovariance2016}.

Implementing state-estimation methods such as Kalman filters and
particle filters requires system-specific knowledge of the physical
properties of the sensor-instrumented system~\cite{barfoot2017state}.
Their implementations also typically require system-specific knowledge
of the constraints imposed by physics on sensor signals.  When
implementing state estimation methods on resource-constrained
embedded systems, this knowledge of the physics of the system is
however just one part of the challenge: System designers must combine
their physical understanding with efficient implementations of the
core linear-algebraic methods or nonlinear dynamical systems, often
in a low-level language such as C, for a hardware platform that has
only tens or hundreds of kilobytes of memory and thus may not be
able to host common libraries such as Eigen~\cite{guennebaud2010eigen}.
As a result, state estimation algorithms for resource-constrained
embedded systems are challenging to implement correctly, challenging
to implement efficiently, and even more challenging to implement
quickly. And, because the implementations of state estimation
algorithms can depend on the physical properties of a system and
its sensors in complex ways, small changes to
the underlying assumptions of the state estimation algorithms can
require significant changes to the state estimation algorithms and
their implementation.

\subsection{Physics-Derived State Estimator Synthesis}
\label{section:introduction:idea}
The behavior of physical systems deviates from the intended behavior with time (\textit{process noise}) 
and measurements of signals which influence the state of a
system will always have some degree of measurement uncertainty
(\textit{measurement noise}). State estimation methods estimate the
future behavior or state of a system in the presence of process
noise and measurement noise. In the absence of process noise, a
fixed time-independent model of a system's behavior would suffice.
Similarly, in the absence of measurement noise and given a model
of system state as a function of external signals, one could
estimate state perfectly.  State estimation methods use a combination
of models of the dynamics of a system and models of the noise in
the system to calculate improved estimates of state variables in
the presence of process noise and measurement noise.

\textbf{Key insight:}\quad The key insight of this work is that we
can use a succinct description of the physical and noise properties
of a system, specified in a recently-developed physics description
language~\cite{lim2018newton}, combined with methods recently-developed
in the machine learning community for Automatic Differentiation of
arithmetic expressions from representations of their abstract syntax
trees (ASTs)~\cite{baydin2017automatic, margossian2019review},
within a compiler.

\subsection{Contributions}
This article makes three key contributions to the state of the art
in state estimation, code generation for embedded systems, and
program synthesis. We present:
\begin{dingautolist}{202}
	\item   \textbf{A new method for automated extended Kalman
			filter (EKF) synthesis}, with and without
			Automatic Differentiation (Section~\ref{section:core}).

	\item   \textbf{An implementation of the method, including
			employing Automatic Differentiation within
			the synthesized state estimators}, for
			extended Backus-Naur form
			(EBNF)~\cite{10.1145/359863.359883}
			expressions (Section~\ref{section:implementation}).

	\item   \textbf{An evaluation of the static and runtime
			behavior of the EKF state estimators
			that our method synthesizes} and evaluation
			of the benefits of Automatic Differentiation
			across embedded processor architectures 
			with and with hardware support for multiplication 
			and with and without support for
			single- and double-precision floating-point
			arithmetic, based on the RISC-V architecture
			(Section~\ref{section:evaluation}).
\end{dingautolist}

\section{Mathematical Foundation}
\label{sec:foundation}
The three central concepts in state estimation are that of a system's
\textit{state}, the system's dynamics or \textit{process}, and
\textit{measurements}.

For example, for an unmanned aerial vehicle (UAV), the state might
comprise the pitch, roll, yaw, position in three-dimensional space
and velocity, as
Figure~\ref{fig:introduction:applications-of-state-estimation},
shown previously, illustrates. The process of the UAV represents
the underlying dynamics of a system such as how its state variables
vary with time. Similarly, the measurements in the UAV example are
the acceleration of the UAV, measured for example using MEMS
accelerometer integrated circuits integrated into the UAV hardware,
or the angular rotational rate, measured for example using MEMS
gyroscope integrated circuits integrated into the UAV hardware.
Even if the UAV moves according to some pre-defined flight path,
its actual flight path (the process) will differ from its planned
flight path. In practice, the UAV uses measurements from its sensors
to estimate its current process state (pitch, roll, yaw, location,
and velocity), but doing so in the presence of deviations from its
planned flight path (e.g., due to cross winds) and in the presence
of noisy readings from its accelerometer and gyro sensors is
challenging.

The \kalman filter~\cite{kalman19601960} is one method to estimate
the state of a dynamic system from noisy measurements of signals
that are related to the state being estimated. It works by iterating
between making predictions based on current estimates and updating
current estimates based on new (noisy) measurements.  This iterative
nature of its equations allows it to be deployed in real-time systems,
continuously integrating new measurements.

\subsection{Notation}
\label{section:foundation:notation}
We notate matrices in an uppercase math boldface font (e.g., matrix
$\matrixSymbol{H}{}$) and vectors in a lowercase math boldface font
(e.g., $\vectorSymbol{v}{}$). We notate all scalar variables in a
standard math italic font (e.g., $k$) and scalar constants (e.g.,
cardinalities) in an uppercase standard math italic font (e.g.,
$N$). We notate estimates with a cap (e.g., $\vectorSymbol{\hat{s}}{k}$)
and the prior, previously-known, or \textit{a priori} assumptions with
a $-$ superscript (e.g., $\vectorSymbol{\hat{s}}{k}^-$).

Let the state of the system being modeled be denoted with vector
$\vectorSymbol{s}{}$ of cardinality $N$ and let $k$ be an index in
time.  Over time, the state of a system will invariably change
across time steps and we define $\vectorSymbol{s}{k}$ as the state
vector at time index $k$.  Let $\matrixSymbol{F}{}$ be the $N \times
N$ \textit{state transition matrix}, a matrix that describes the
transformations of state across time steps.

Many systems have an ability to influence their state (e.g., by
controlling motors that determine their propulsion).  Let vector
$\vectorSymbol{u}{k}$ be the vector of controllable parameters of
the system, with cardinality $U$, at time step $k$. Knowledge of
these controllable parameters is useful in state estimation if
available, but is not essential to most state estimation methods.
Let $\matrixSymbol{B}{}$ be the $N \times U$ matrix that scales the
elements of the \textit{controllable parameter vector} or \textit{control
vector} $\vectorSymbol{u}{k}$.  In a system for which these
controllable parameters are not known, $\matrixSymbol{B}{}$ is not
used. For example, in a land vehicle where we can measure the angular
rate of wheels but where we cannot
measure the linear velocity of the vehicle, the matrix $\matrixSymbol{B}{}$
would be the transform mapping from wheel rotation rate to velocity.

Let $\vectorSymbol{n}{p,k}$ be the noise vector with cardinality
$N$ (in a system with $N$ process states) in a process at time step
$k$ and let $\vectorSymbol{n}{m,k}$ be the noise vector of cardinality
$Z$ (in a system with $Z$ sensors), at time step $k$.  Let
$\vectorSymbol{z}{k}$ be the vector of cardinality $Z$ comprising
the estimates of the true underlying signals being measured by
sensors, based on the current state and the assumption of
the measurement noise.  If there was no noise term in
Equation~\ref{eq:measure}, then one could estimate the state
$\vectorSymbol{s}{k}$ directly as the product of the inverse of
$\matrixSymbol{H}{}$ and the measurement $\vectorSymbol{z}{k}$.

In this work, we derive $\matrixSymbol{F}{}$ and $\matrixSymbol{H}{}$
of Equations~\ref{eq:predict_state} and \ref{eq:predict_cov}
from a machine-readable description of the system.

\subsection{Linear Kalman Filter}
\label{section:lkf_input}
The simplest form of the \kalman filter is the linear \kalman filter (LKF),
where the dynamic process and measurement functions are described
by linear equations. A key insight of the LKF is that
it incorporates prior information (a model) about the properties
of the physical system that state variables represent. This prior
information on the state vector and control vector enable the \kalman
filter to predict both the state and the inter-state-variable
covariance while capturing the uncertainty.

Let $\matrixSymbol{H}{}$ be the $Z \times N$ measurement matrix,
let $\matrixSymbol{P}{}$ be the $N \times N$
state estimate covariance matrix, let $\matrixSymbol{Q}{}$ be the
$N \times N$ process uncertainty matrix, let $\matrixSymbol{R}{}$
be the $Z \times Z$ measurement uncertainty matrix, and let
$\matrixSymbol{K}{}$ be an $N \times Z$ matrix.

\noindent\textbf{Intuition}: $\matrixSymbol{K}{}$ represents the
comparison between the current covariance and the measurement noise
matrix. The smaller the entries in the process uncertainty matrix
$\matrixSymbol{Q}{}$, the more the filter favors the current state.
The larger the covariances in the matrix $\matrixSymbol{P}{}$ and
the smaller the measurement uncertainty (entries in the matrix
$\matrixSymbol{R}{}$), the more the \kalman filter favors the
measurements. In the limit as the entries in the covariance matrix 
$\matrixSymbol{P}{}$ approach zero, the \kalman gain matrix $\matrixSymbol{K}{}$
approaches $\matrixSymbol{H}{}^{-1}$ therefore, the new state is based only on the prior
of the physics model. Similarly, in the limit when all the entries in
$\matrixSymbol{R}{}$ approach 0, the new state will be based entirely
on measurements.

\begin{table}%
\caption{Linear Kalman filter equations. The method we present in
Section~\ref{section:core} extracts the information in the terms
related to the \highlightPhysics{\mbox{physics}} of the system as
well as properties of the \highlightNoise{\mbox{noise}} when provided
in the input to the method.}
\begin{tabular}{rm{7.0cm}}
\toprule
\textbf{Process}: & \begin{equation} \label{eq:process} \vectorSymbol{s}{k} = \highlightPhysics{\matrixSymbol{F}{}\vectorSymbol{s}{k-1} + \matrixSymbol{B}{}\vectorSymbol{u}{k}} + \highlightNoise{\vectorSymbol{n}{p,k}} . \end{equation} \\[-4ex]
\textbf{Measure}: & \begin{equation} \label{eq:measure} \vectorSymbol{z}{k} = \highlightPhysics{\matrixSymbol{H}{}\vectorSymbol{s}{k}} + \highlightNoise{\vectorSymbol{n}{m,k}} . \end{equation} \\[-4ex]
 & \\
\midrule
\textbf{Predict}: & \begin{equation} \vectorSymbol{\hat{s}}{k}^- = \matrixSymbol{F}{}\vectorSymbol{\hat{s}}{{k-1}} + \matrixSymbol{B}{}\vectorSymbol{u}{k} . \label{eq:predict_state} \end{equation}\\[-4ex]
         & \begin{equation} \matrixSymbol{P}{k}^- = \matrixSymbol{F}{}\matrixSymbol{P}{k-1}\matrixSymbol{F}{}^{\intercal} + \matrixSymbol{Q}{} \label{eq:predict_cov} . \end{equation}\\[-4ex]
 & \\
\textbf{Gain}:   & \begin{equation} \matrixSymbol{K}{k} = \matrixSymbol{P}{k}^-\matrixSymbol{H}{}^\intercal(\matrixSymbol{H}{}\matrixSymbol{P}{k}^-\matrixSymbol{H}{}^\intercal + \matrixSymbol{R}{})^{-1} . \label{eq:kalman_gain} \end{equation}\\[-4ex]
\textbf{Update}: & \begin{equation} \vectorSymbol{\hat{s}}{k} = \vectorSymbol{\hat{s}}{k}^- + \matrixSymbol{K}{k}(\vectorSymbol{z}{k} - \matrixSymbol{H}{}\vectorSymbol{\hat{s}}{k}^-) . \label{eq:update_state} \end{equation}\\[-4ex]
        & \begin{equation} \matrixSymbol{P}{k} = (\matrixSymbol{I}{} - \matrixSymbol{K}{k}\matrixSymbol{H}{})\matrixSymbol{P}{k}^- . \label{eq:update_cov} \end{equation}\\[-4ex]
\bottomrule
\end{tabular}
\label{tab:linearKalmanFilterEqs}
\end{table}

Table~\ref{tab:linearKalmanFilterEqs} summarizes the equations of
the linear \kalman filter.  The equations in
Table~\ref{tab:linearKalmanFilterEqs} show the simple idea of
maintaining a state estimate, updating it on new measurements
according to a gain $\matrixSymbol{K}{k}$, that is determined by
comparing our faith in our current estimates with the uncertainty
of the measurements.  Equations~\ref{eq:process} and~\ref{eq:measure}
correspond to the noisy process and measurement models of the system.
The terms $\vectorSymbol{n}{p,k}$ and $\vectorSymbol{n}{m,k}$ are
the zero-mean white noise for the process and the measurement model,
respectively.  Equations~\ref{eq:predict_state} and~\ref{eq:predict_cov}
formulate the prediction step, in which we propagate our current
estimates of the state and its covariance through the process model.
Equations~\ref{eq:update_state} and~\ref{eq:update_cov} update the
state and its covariance based on new measurements and the \kalman
gain, calculated using Equation~\ref{eq:kalman_gain}.

\textbf{Input of the \kalman filter:} In a linear \kalman
filter the state-space is usually represented as a set of matrices,
with $N$, $Z$, and $U$ being the dimensions of the state vector,
measurement vector, and input vector, respectively.  The inputs of
the system in this case are:
\begin{dingautolist}{192}
    \item $\matrixSymbol{F}{}$: State transition $N \times N$ matrix.
    \item $\matrixSymbol{H}{}$: Measurement $Z \times N$ matrix.
    \item $\vectorSymbol{s}{0}$: Initial state estimate $N \times 1$ vector.
    \item $\matrixSymbol{P}{0}$: Initial state estimate covariance $N \times N$ matrix.
    \item $\matrixSymbol{B}{}$: Control transition $N \times U$ matrix.
    \item $\matrixSymbol{Q}{}$: Process uncertainty $N \times N$ matrix.
    \item $\matrixSymbol{R}{}$: Measurement uncertainty $Z \times Z$ matrix.
\end{dingautolist}
Of these, $\matrixSymbol{F}{}$ and $\matrixSymbol{H}{}$ are crucial
to describing the model of the system because the filter equations
need a way to propagate the state (achieved via $\matrixSymbol{F}{}$)
and to correlate sensor readings with the predicted state (achieved
via $\matrixSymbol{H}{}$).  
The process and measurement noise may be known from
sensor descriptions or can be inferred using auto-covariance methods
on sample data~\cite{odelson2006new}.  The drawback of automatic
inference is that it is usually less accurate compared to the case
where the sensor noise distributions are provided.

\subsection{Extended Kalman Filter}
\label{subsec:EKF}
Perhaps the most widely known variant of the \kalman filter is the
Extended \kalman filter (EKF), derived in an effort to extend the
original linear equations for usage in non-linear systems.  In EKF,
the new state vector and the measurement vector result from the old
state by the application of functions $\functionSymbol{f}$ and
$\functionSymbol{h}$ instead of a matrix multiplication.

These functions can be non-linear, such as transcendental functions
(e.g., $\sin$ and $\cos$), which are very often useful when describing
physical systems.  The application of these functions mandates the
adaptation of the the \textit{predict} and \textit{update} equations.
The updated equations are summarized in
Table~\ref{tab:extendedKalmanFilterEqs}.  For the rest of the
equations the $\matrixSymbol{F}{}$ and $\matrixSymbol{H}{}$ matrices
now represent the Jacobian of the functions $\functionSymbol{f}$
and $\functionSymbol{h}$ at the current state with respect to the
state vector variables.

\begin{table}%
\caption{Extended Kalman filter equations. The Automatic Differentiation method
presented in Section~\ref{section:autodiff_ekf} automates the
process of determining the Jacobians of the functions $\functionSymbol{f}$
and $\functionSymbol{h}$.}
\begin{tabular}{rm{7.0cm}}
\toprule
\textbf{Process}: & \begin{equation} \label{eq:goal} \vectorSymbol{s}{k} = \highlightPhysics{\functionSymbol{f}(\vectorSymbol{s}{k-1}, \vectorSymbol{u}{k})} + \highlightNoise{\vectorSymbol{n}{p,k}} . \end{equation} \\[-4ex]
\textbf{Measure}: & \begin{equation} \label{eq:goal2} \vectorSymbol{z}{k} = \highlightPhysics{\functionSymbol{h}(\vectorSymbol{s}{k})} + \highlightNoise{\vectorSymbol{n}{m,k}} . \end{equation} \\[-4ex]
 & \\
\midrule
\textbf{Predict}: & \begin{equation} \vectorSymbol{\hat{s}}{k}^- = \functionSymbol{f}(\vectorSymbol{\hat{s}}{k-1}, \vectorSymbol{u}{k}) \label{eq:ekf_predict} . \end{equation}\\[-4ex]
\textbf{Update}: & \begin{equation} \vectorSymbol{\hat{s}}{k} = \vectorSymbol{\hat{s}}{k}^- + \matrixSymbol{K}{k}(\vectorSymbol{z}{k} - \functionSymbol{h}(\vectorSymbol{\hat{s}}{k}^-)) . \label{eq:ekf_update} \end{equation}\\[-3ex]
\bottomrule
\end{tabular}
\label{tab:extendedKalmanFilterEqs}
\end{table}

\textbf{Input of the Extended \kalman filter: }
Similar to the LKF, $N$ is the dimension of the state vector,
$Z$ is the dimension of the measurement vector,
$U$ is the dimension of the input vector,
while $dt$ denotes the time difference between two successive time transitions.
Additionally, we require information on the additivity of the
model uncertainties $\matrixSymbol{Q}{}$ and $\matrixSymbol{R}{}$.
The inputs in EKF case are:
\smallskip
\begin{dingautolist}{192}
    \item $\functionSymbol{f}$: State transition function of type $(N, U, dt)\rightarrow N$.
    \item $\functionSymbol{h}$: Measurement function of type $N \rightarrow Z$.
    \item $\vectorSymbol{s}{0}$: Initial state estimate $N \times 1$ vector.
    \item $\matrixSymbol{P}{0}$: Initial state estimate covariance $N \times N$.
    \item $\matrixSymbol{Q}{}$: Process uncertainty $N \times N$ matrix.
    \item $\matrixSymbol{R}{}$: Measurement uncertainty $Z \times Z$ matrix.
    \item Boolean additivity values for $\matrixSymbol{Q}{}$ and $\matrixSymbol{R}{}$. %
\end{dingautolist}
\smallskip
As in the LKF case, $\functionSymbol{f}$ and $\functionSymbol{h}$ are crucial to describing the model
of the system.

The EKF is the \textit{de facto} standard for
state estimation in non-linear systems due to its theoretical
simplicity and ease of implementation compared to other
approaches~\cite{wan2000unscented}.  Despite its popularity, it
suffers from two fundamental drawbacks~\cite{haykin2009cubature}.
Firstly, it uses a first-order Taylor series expansion to locally
linearize the non-linear model, and thus requires the non-linearities
to be mild.  Secondly, its employment requires the derivation of
the Jacobian matrices of the non-linear dynamic models of the system,
which, if exist at all~\cite{julier2004unscented}, are highly 
impractical to compute.
We overcome this impracticality, using Automatic
Differentiation~\cite{baydin2017automatic, margossian2019review}
as we discuss in Section~\ref{section:autodiff_ekf}.

\section{Automated Physics-Derived State Estimator Synthesis}
\label{section:core}

Linear \kalman filters (LKFs) and extended \kalman filters (EKFs)
are often employed as the state estimation algorithms in cyber-physical
or embedded systems. As detailed in
Section~\ref{sec:foundation}, state estimation methods such as the
LKF and EKF require an understanding and model of the dynamics of
the physical system for which they are implemented. This makes the
rapid implementation of state estimation and sensor fusion methods
challenging, especially on resource-constrained embedded systems
where both execution time and memory are at a premium. As a result,
there is an unmet need for automated methods for synthesizing
small-footprint implementations of state estimation such as LKFs
and EKFs, in low-level languages such as C, from high-level
descriptions of their measurement and process dynamics.

Recent research has shown increasing interest in the development
of specification languages that allow the users to specify properties
of physical systems systems~\cite{lim2018newton, wang2019deriving}.
We extend the specification capabilities of the Newton
language~\cite{lim2018newton} to allow users to input the necessary
information for the generation of \kalman filters and we implement
a code generation methodology as a new backend for the Newton
compiler. The resulting system provides an automated method for
mapping high-level system descriptions to low-level source code
that can be integrated into an embedded system.

\begin{figure}
  \centering
  \begin{lstlisting}[language=Newton]
  include "NewtonBaseSignals.nt"
  
  g : constant = 9.80665 ajf;
  
  # Ideal pendulum
  pendulum_process : invariant( theta	: angle,
                    dtheta	: angularRate,
                    dt		: time,
                    L		: distance) =
  {
    theta ~ theta + dtheta * dt,
    dtheta ~ dtheta - g/L * sin(theta) * dt
  }
  
  
  pendulum_measure : invariant( theta	: angle,
                    dtheta	: angularRate,
                    dt		: time,
                    gyro_z	: angularRate ) =
  {
    gyro_z ~ dtheta
  }
  \end{lstlisting}
  \caption[LoF entry]{Physical description equations for the process
  and measurement models written in Newton~\protect\cite{lim2018newton}.
  The pendulum process invariant is modeled after the ideal pendulum 
  (Equation~\ref{eq:PEND_1}), derived as explained in Section~\ref{subsec:example}.
  The invariants' arguments list all identifiers involved in the equations, and 
  their respective dimensions. 
  Any identifier that is not a state or a measurement variable will 
  be passed on to the generated functions as an extra argument.
  \label{fig:models-example}}
\end{figure}

\subsection{Input Specifications}
\label{subsec:input}
The researcher or engineer using our state estimation synthesis
methodology requires a method of providing an input description of
a \kalman filter according to the parameters and equations of
Section~\ref{sec:foundation}.  We provide a flexible and general
way of representing this information, by extending the capabilities
of the Newton compiler to parse equations that describe the target
physical system.  This enables a unified approach for describing
both linear and non-linear models and additionally allows a user
of the system to define the noise distribution of the input functions.

\subsection{Example: Pendulum}
\label{subsec:example}
As a concrete example, we examine the case of the
state estimation of a vertical ideal pendulum.
\textbf{Notation:} Let $\theta$ be the
angle of displacement from the resting point, $g$ the
acceleration due to gravity, $l$ the length of the rod, $b$ the damping constant 
with units of $kg\,s^{-1}$, $m$ the mass of the pendulum's bob, 
and $t$ the the elapsed time. Then, Equation~\ref{eq:PEND_1}
describes the dynamics of the oscillation:

\begin{equation}
    \dv[2]{\theta}{t} + \frac{b}{m}\dv{\theta}{t} + \frac{g}{l}\sin(\theta) = 0 .
    \label{eq:PEND_1}
  \end{equation}

\noindent\textbf{State in the pendulum example:} We define the state
of the example system as the vector of the angle $\theta$ and the
derivative $\dv{\theta}{t}$.  Our goal is to generate a
\kalman filter which tracks the state according to the measurements
from a gyroscope attached to the mass of the pendulum. To achieve
that, the user must specify the \textit{process} and \textit{measurement}
models of the pendulum.

\noindent\textbf{Input specification in the pendulum example:}
Figure~\ref{fig:models-example} shows the Newton description of the
equations for the \textit{process} and \textit{measurement} models
in the example system.  The Newton language provides the \code{invariant}
keyword to declare a description of the physical system, whose form
remains invariant throughout the lifetime of the physical object.
The \code{\textasciitilde} Newton language comparison operator
allows a clause in an invariant to specify that the left and
right hand sides of the operator have the same units of measure and
are proportional.

The \textit{process} model of the pendulum (\code{pendulum\_process})
appears in a linearized form, split over two equations, in order
to track $\theta$ and its derivative separately.  The first equation
is a time transition for $\theta$, which changes by $\dv{\theta}{t}dt$.
The second equation is a time transition for $\dv{\theta}{t}$, which
changes by $\dv[2]{\theta}{t}$, according to Equation~\ref{eq:PEND_1}.
The \textit{measure} invariant (\code{pendulum\_measure}) corresponds
to the z axis of the sensor data captured by the gyroscope.  This
measurement corresponds to the angular rate of the pendulum, i.e.,
$\dv{\theta}{t}$.

In the general case, we make the following assumptions regarding
the input description of the target physical system:
\begin{enumerate}
	\item The identifiers of the \textit{process} model invariant
	and the \textit{measurement} model invariant are specified
	in the input.

	\item All the left-values of the \textit{process} model
  invariant constraints are part of the state of the system and
  all the left-values of the \textit{measurement} model
  invariant constraints are measurement variables.
\end{enumerate}

\subsection{Code Generation Algorithm}
\label{subsec:codeGenAlgo}
In this step, the user-provided description of the target physical
system must be processed in order to generate the source code of a
\kalman filter.  This code is composed of the following three
functions which are essential for the correct operation of the
filter:
\begin{itemize}
\item \textbf{\code{filterInit()}:} Initializes state and state covariance.
\item \textbf{\code{filterUpdate()}:} Ingests the input sensor measurements and incorporates them to the current state and covariance.
\item \textbf{\code{filterPredict()}:} Projects the state in time according to the \textit{process} model.
\end{itemize}

The user input Newton description is parsed by the Newton parser
and converted into an abstract syntax tree (AST).  This AST is the
input to Algorithm~\ref{algo-kalman-synthesis}, the \kalman filter 
code generation algorithm.

First, from the AST, the algorithm identifies the invariants that
describe the \textit{process} and \textit{measurement} models (Step
1). Next (Step 2), the algorithm differentiates between variables
that belong to the state versus those that belong to the measurement,
as well as any extra variables that must be passed to the filter
functions, according to the input specification.  Next (Step 3),
the algorithm generates the required data structures for the filter
operation.  This includes C language \code{struct} definitions for
the core data of the filter (e.g., the current state and covariance),
as well as enumerations of constant values.  Based on the information
gathered by the three initial steps, the algorithm identifies which
structures and variables need to be initialized before the execution
of the filter and thus produces the \code{filterInit()} function
in Step 4.

Next (Step 5) the algorithm performs the operations for
generating the source code of the \code{filterPredict()} function.
The algorithm performs a linearity check in Step 5a, by examining
the factors of the state variables and the functions used in the
expressions.  If it detects a non-linear case, it generates the
required C expressions in Step 5b.  In a linear case Step 5b is skipped.
In Step 5c the algorithm generates the propagation
matrix for the linear case or the Jacobian for the non-linear case.
In Step 5d the algorithm generates the body of the \kalman\ filter
\code{filterPredict()} and it connects the generated function to
the results of the previous steps.

The algorithm follows a similar code generation flow in Step 6, to
generate the source code of the \code{filterUpdate()} function. The
algorithm again performs a linearity check in Step 6a, and accordingly
executes Steps 6b and 6c.  In Step 6d the algorithm generates the
necessary matrix operations for the calculations of the
\code{filterUpdate()} equations of the \kalman\ filter.

\begin{algorithm}[tb]
\footnotesize
\KwData{AST of input Newton description}
\SetKwFunction{OnUAV}{Trajectory calculation}
  \BlankLine

  \textbf{Step 1:} Identify process and measurement model invariants.

  \textbf{Step 2:} Discern state and measurement variables.

  \textbf{Step 3:} Generate data required filter data structures.

  \textbf{Step 4:} Generate \code{filterInit()} function.

  \textbf{Step 5:} Generate \code{filterPredict()} function:

  \Indp\Indp

  \textbf{Step 5a:} Determine linearity.

  \textbf{Step 5b:} Generate state propagation functions.

  \textbf{Step 5c:} Construct state propagation or Jacobian ($\matrixSymbol{F}{}$).

  \textbf{Step 5d:} Generate predict step matrix operations.

  \BlankLine

  \Indm\Indm

  \textbf{Step 6:} Generate \code{filterUpdate()} function:

  \Indp\Indp

  \textbf{Step 6a:} Determine linearity.

  \textbf{Step 6b:} Generate state measurement functions.

  \textbf{Step 6c:} Construct state measurement or Jacobian ($\matrixSymbol{H}{}$).

  \textbf{Step 6d:} Generate update step matrix operations.

  \caption{State estimation generation algorithm}
  \label{algo-kalman-synthesis}
\end{algorithm}

\subsection{Automatic Differentiation for the EKF}
\label{section:autodiff_ekf}

As mentioned in Section~\ref{subsec:EKF}, a considerable obstacle
in the incorporation of the EKF in embedded systems arises from the
need to derive the Jacobian matrices of the non-linear dynamic
models of the system. We address this
challenge by using Automatic Differentiation (AutoDiff), a
technique that computes the derivatives of code operation using the
chain rule of derivatives.

AutoDiff is neither a numerical approximation of the derivative nor
a symbolic method for deriving it.  It can be considered as a
non-standard interpretation of a computer program where the output
is an augmented computer-program that also contains the calculations
for various derivatives.
AutoDiff has two modes of operation: \textit{forward mode} and \textit{reverse mode}.
Consider the expression in Equation~\ref{eq:original}
(Table~\ref{tab:linearKalmanFilterEqs2}) in order to explain each
mode.
The preliminary step is to express the expression in Equation~\ref{eq:original}
in a static single assignment (SSA) form~\cite{alpern1988detecting, braun2013simple} as presented in 
Equations~\ref{eq:SSA_1} to~\ref{eq:SSA_4}.

\noindent\textbf{Forward mode:} In this mode the intermediate
derivative of an expression with respect to one of the input variables
can be calculated alongside the normal function calculation.  We provide
an example by showing the steps required to differentiate the
expression in Equation~\ref{eq:original} with respect to an undefined
variable $t$ (Table~\ref{table:autoDiffModes}). We first produce the SSA form of the expression, as
shown Equations~\ref{eq:SSA_1} to~\ref{eq:SSA_4}.
By setting $t=x$ we can calculate $\pdv{z}{x}$, assuming $\pdv{y}{x}=0$.
Similarly, we can calculate $\pdv{z}{y}$ by setting $t=y$.  In
general, $n$ partial derivatives can be calculated with $n$ separate
function evaluations.  A small overhead incurs because the function
is extended to calculate the derivatives alongside execution.  The
overhead for the Forward Mode is approximately $O(n)$, where $n$ is the
number of operations in the original function.

\begin{table}%
  \caption{Example \highlightPreSSA{\mbox{expression}} and its \highlightPostSSA{\mbox{ssa form code sequence}}.}
  \begin{tabular}{C{4.0cm}|C{4.0cm}}
  \toprule
  \centering \textbf{Original} & \textbf{SSA Form}\\
  \midrule\\[-4ex]
  \begin{equation} \highlightPreSSA{z = x*y^2 + \sin(x).} \label{eq:original} \end{equation} & \begin{equation} \highlightPostSSA{r_1 = y^2;}  \label{eq:SSA_1} \end{equation}\\[-8ex]
          & \begin{equation} \highlightPostSSA{r_2 = x \cdot r_1;}  \end{equation} \\[-8ex]
          & \begin{equation} \highlightPostSSA{r_3 = \sin(x);}  \label{eq:SSA_3} \end{equation}\\[-8ex]
          & \begin{equation} \highlightPostSSA{z = r_2 + r_3.} \label{eq:SSA_4} \end{equation}\\
  \bottomrule
  \end{tabular}
  \label{tab:linearKalmanFilterEqs2}
\end{table}

\begin{table}[b]
  \caption{Modes for Automatic Differentiation.}
  \begin{tabular}{C{3.6cm}|C{4.4cm}}
  \toprule
  \centering \textbf{Forward Mode} & \textbf{Reverse Mode} \\
  \midrule
  \begin{equation} \pdv{r_1}{t} = 2\cdot\pdv{y}{t} \label{eq:FM_1} \end{equation} & \begin{equation} \pdv{s}{r_3} = \pdv{s}{z} \label{eq:RM_1} \end{equation}\\[-5ex]
  \begin{equation} \pdv{r_2}{t} = x \cdot \pdv{r_1}{t} + r_1 \cdot \pdv{x}{t} \end{equation} & \begin{equation} \pdv{s}{r_2} = \pdv{s}{z} \end{equation}\\[-5ex]
  \begin{equation} \pdv{r_3}{t} = \cos(x) \cdot \pdv{x}{t} \end{equation} & \begin{equation} \pdv{s}{r_1} = x \cdot \pdv{s}{r_2} \end{equation}\\[-5ex]
  \begin{equation} \pdv{z}{t} = \pdv{r_2}{t} + \pdv{r_3}{t} \label{eq:FM_4} \end{equation} & \begin{equation} \pdv{s}{x} = r_1 \cdot \pdv{s}{r_2} + \cos(x) \cdot \pdv{s}{r_3} \end{equation}\\[-5ex]
   & \begin{equation} \pdv{s}{y} = 2 \cdot \pdv{s}{r_1} \label{eq:RM_5} \end{equation}\\
  \bottomrule
  \end{tabular}
  \label{table:autoDiffModes}
\end{table}

\noindent\textbf{Reverse mode:} In this mode, the calculation of
the output of the function takes place first, untangled from the
calculations for the partial derivatives.  Afterwards, the partial
derivatives of the output with respect to each intermediate result
are calculated downwards, in a fountain-like manner, until those
with respect to each input variable are reached.
Starting from Equation~\ref{eq:SSA_4}, the final assignment of the
SSA form of the expression in Equation~\ref{eq:original} we
differentiate an undefined variable $s$ with
respect to each one of the left values in Equations~\ref{eq:SSA_1}
to~\ref{eq:SSA_3}, i.e., $r_1$ to $r_3$.  This can be perceived as
finding out what output variables a given input variable can affect
and by how much.  The updated expressions are provided in Table~\ref{table:autoDiffModes}.
By setting $s = z$ we get both $\pdv{z}{x}$ and $\pdv{z}{y}$ in the same run.
As a result, $n$ partial derivatives can be calculated with one function evaluation
and a small overhead due to the computation of the partial
derivatives. The computational overhead, similarly to Forward Mode, is roughly $O(n)$, but
Reverse Mode needs memory space proportional to the complexity of the original function 
for storing intermediate values. 
The Forward Mode AutoDiff is able to calculate the partial derivatives of all the output variables with respect
to one input variable in the same pass, whereas in Reverse Mode AutoDiff we can calculate the partial derivatives
of one output variable with respect to all input variables in the same pass.
In the context of calculating the Jacobian, Forward Mode computes columns while Reverse Mode computes rows.
As such, with AutoDiff, a precise calculation of the Jacobian can be performed in $n$ evaluations
of the original function~\cite{baydin2017automatic, margossian2019review}. 
In our case the user inputs the model as a series of equations, one for each output variable. 
For this, it is more convenient to implement the Reverse Mode.

The AutoDiff engine was implemented as compiler back-end and uses a three-pass method 
to annotate the AST nodes and then generate the output code.
In the first pass, with a pre-order traversal, each node of the tree is assigned
a ``variable identifier''.
This will identify the intermediate result of this node in the SSA form.
In the second pass, with a post-order traversal, the SSA form of the expression
is printed to a buffer using those identifiers.
In the third, with a pre-order traversal, the reverse SSA form for the
calculation of the chained partial derivatives is exported.

\subsection{End-to-End Example}
\begin{figure*}
	\centering
	\includegraphics[width=1.0\linewidth]{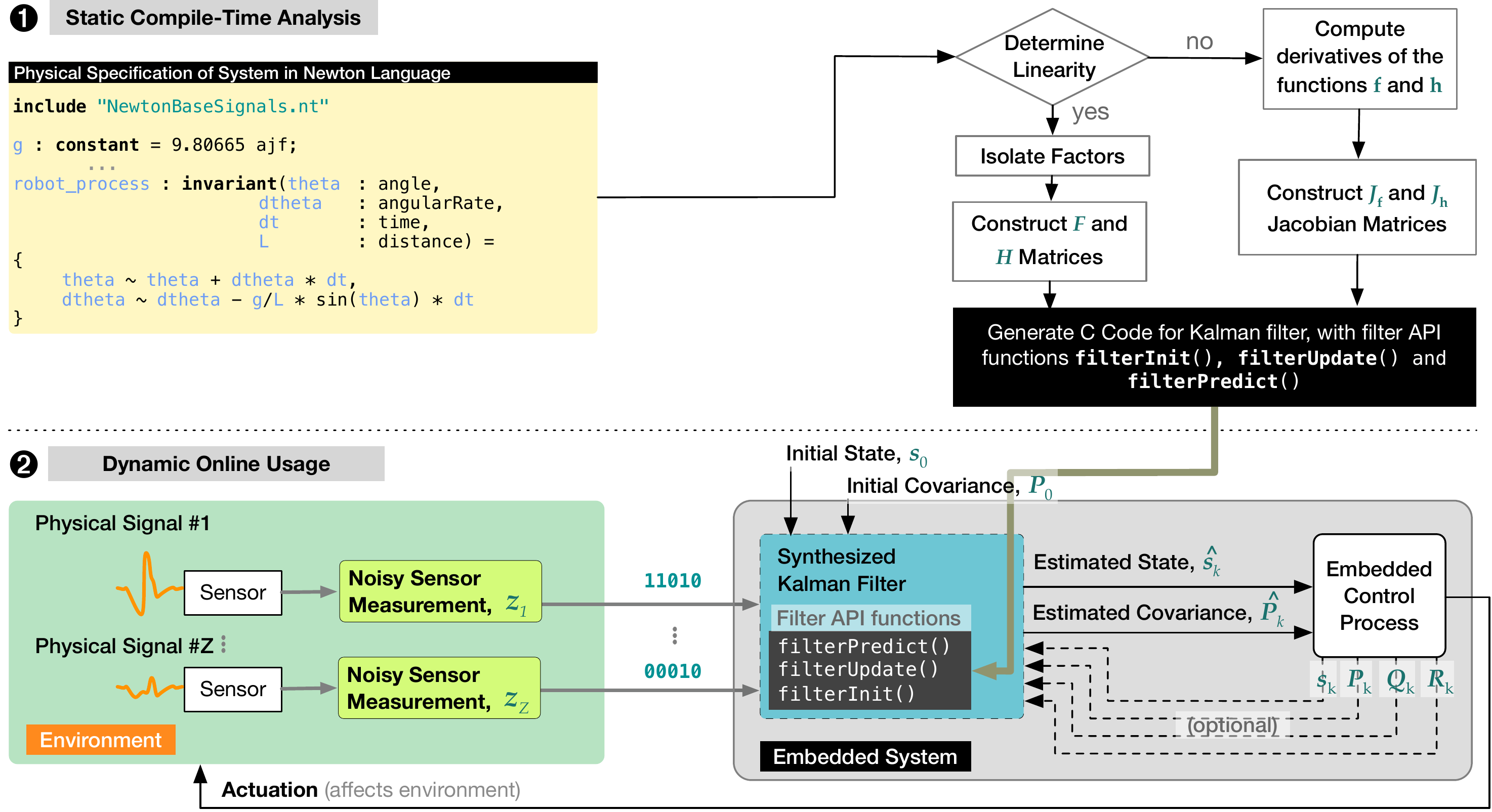}
  \caption{System overview. In the \textbf{Static Compile-Time Analysis} stage the user 
  provides a physical specification of the embedded system in question, encoded as 
  an Newton language description file. The Newton compiler parses and dimensionally checks the file. 
  The generated AST is directed to the code generation back-end which starts with a 
  linearity check and proceeds accordingly to the generation of the expressions and the matrices 
  required by the \kalman\ filter algorithm. The algorithmic steps involved in this process have been presented 
  in Section~\ref{subsec:codeGenAlgo}.
  In the \textbf{Dynamic Online Usage} stage the user evaluates the synthesized \kalman 
  filter either in simulation or in the actual target embedded systems. 
  The run-time stage of the filter commences with the initialization of 
  the state and covariance.
  When the target computing system establishes its interaction with the physical 
  environment, the prediction and update functions of the \kalman filter are 
  invoked according to the target application control flow. 
  The dynamically estimated state and covariance are available to 
  any scope that can access the core data structures of the filter.
  An optional injection of a state is also supported.
	\label{fig:lkf1}}
\end{figure*}

Our code generation process consists of two main stages. \textbf{Static
Compile-Time Analysis} and \textbf{Dynamic Online Usage}.
Figure~\ref{fig:lkf1} illustrates an end-to-end example of the operation and connections of
all the subcomponents presented in Sections~\ref{subsec:input} to~\ref{section:autodiff_ekf}, 
to achieve automated \kalman filter source code generation. 

\section{Implementation}
\label{section:implementation}

\subsection{API of the Interaction with the Synthesized Filter}
\label{sec:synthesized_api}

The output of the synthesis is the generated \kalman filter source code, either LKF or EKF,
which follows the conventions presented in Section~\ref{section:core}.
We instruct the synthesized source code to automatically interface 
the final binary against a lightweight implementation of the necessary linear 
algebra functions. The user also has the option to provide their own libraries.

For the case of EKF, the user can choose between Standard and Automatic
Differentiation. The key differences between these in the generated code are: 
(i) the state and measurement equations use functions instead of matrices; 
(ii) the Jacobian matrix is calculated at each step using the derivatives of state and measurement 
functions as follows.
\noindent\textbf{Standard differentiation:} Another set of $O(N^2+Z^2)$ 
functions are created for the derivative of each function with regards
to each state variable.
\noindent
\textbf{Automatic differentiation:} The generation of the first $O(N+Z)$ 
functions is hi-jacked to produce the SSA form code of 
the expression and the Reverse Mode AutoDiff calculation of the derivatives,
detailed in~\ref{section:autodiff_ekf}.
The code generation back-end
outputs code fragments by traversing the AST corresponding to each user input. 
Reusable intermediate results are kept internally for use at subsequent steps of the 
code generation back-end. 
The source code is incrementally stored in a buffer and eventually written to a user-specified file. %

\subsection{Newton Language: Extensions and Limitations}
Support for the required constructs to model \kalman filters in the Newton
language~\cite{lim2018newton} grammar varies.
Matrix notation is not directly supported in the grammar but expression lists 
inside the invariants are.

Newton implements a dimensional type system that uses signals as types 
(e.g. \textit{angle, time and distance} in Fig.~\ref{fig:models-example}).
Primitive signals (e.g. the S.I. units) are defined in the include file, \textit{NewtonBaseSignals.nt}.
Dimensions can be applied to an identifier through a signal and new signals can 
be declared. A Newton description must satisfy this type system, i.e. be dimensionally correct, 
for the compiler to accept it. This is an additional way to prevent wrong filters 
from being generated.

Uncertainty notation is implemented as an extension of the signal declaration statement. 
Any signal can be optionally enhanced with uncertainty information by providing the expected
distribution and its parameters as an argument. In the context of this work we have only 
worked with Gaussian distributions, but the compiler can be extended in a straightforward
manner to support other distributions.

The engine for Automatic Differentiation (Section~\ref{section:autodiff_ekf}) of 
Newton expressions was implemented as a three-pass algorithm of the compiler AST. 
By design, it is abstracted from the estimator generation back-end, making 
the technique available to any code generation back-end or other compiler with 
similar AST structure.

\section{Experimental Evaluation}
\label{section:evaluation}

\begin{figure*}
    \subfigure[Experiment 1: Estimation of oscillation angle over time.]{ %
    \centering
    \includegraphics[width=0.95\columnwidth]{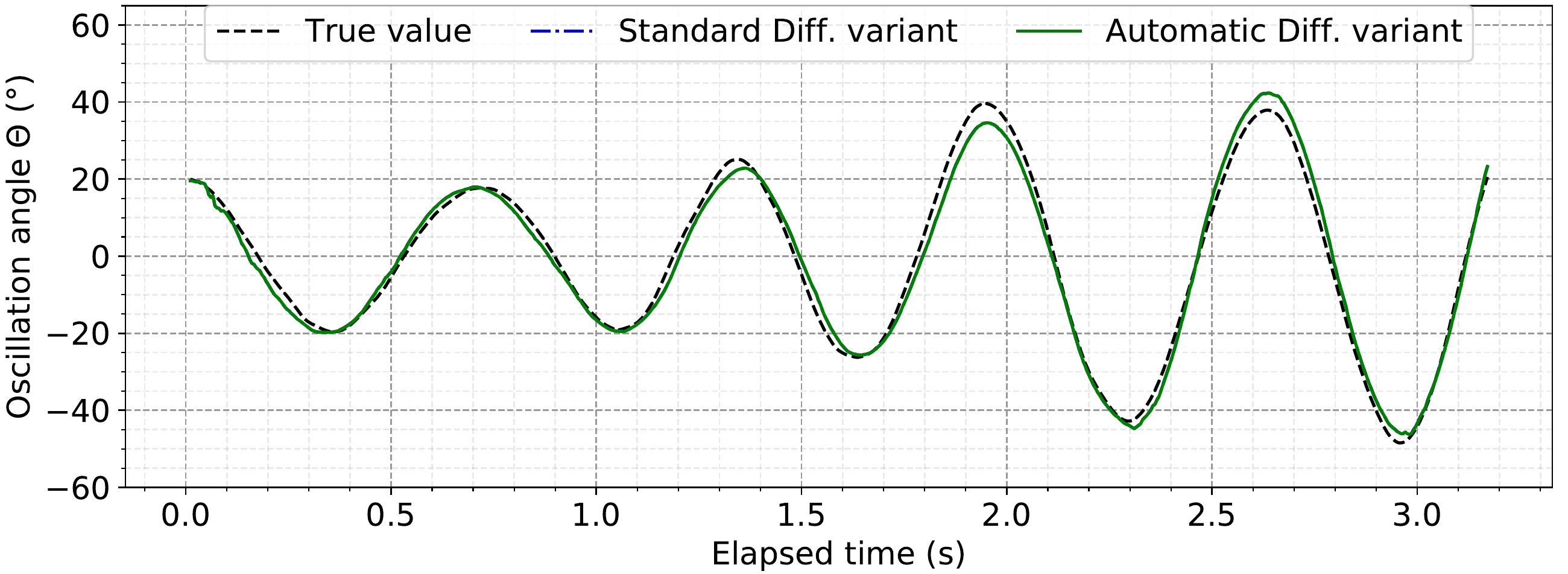}
        \label{fig:pendulum-wo-damp-a}
    }
    \subfigure[Experiment 2: Estimation of oscillation angle over time.]{ %
	\centering
    \includegraphics[width=0.95\columnwidth]{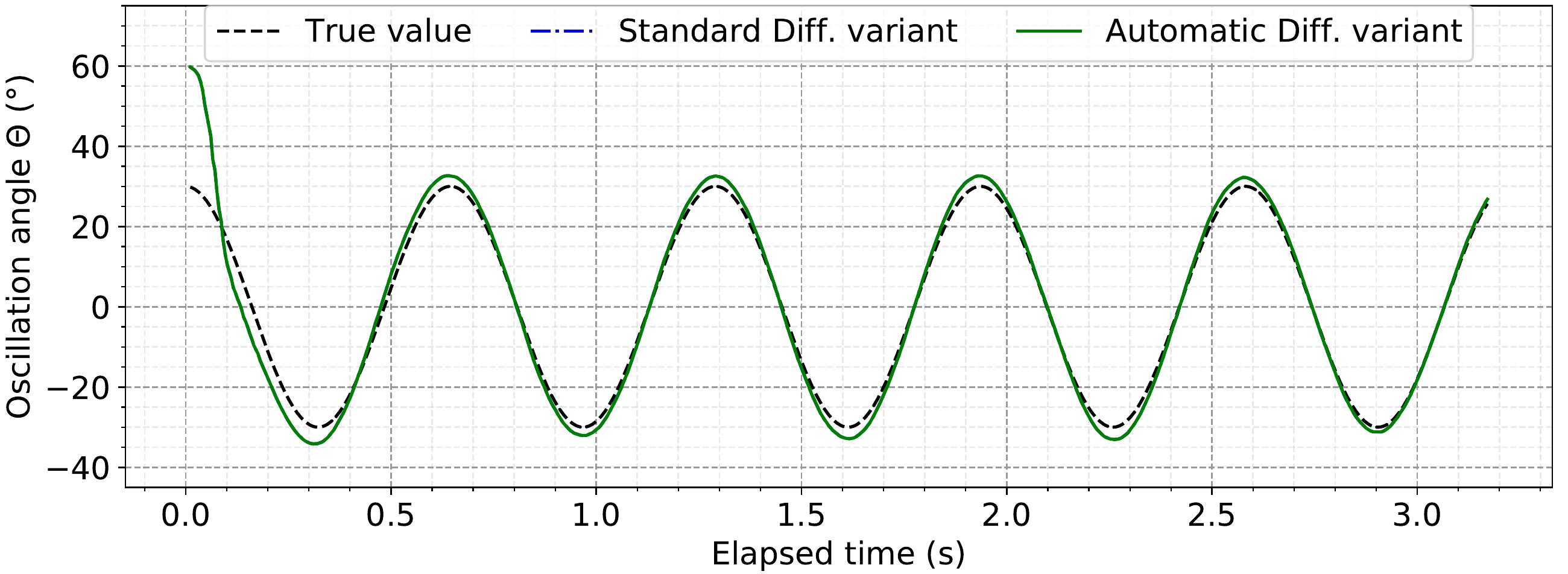}
    \label{fig:pendulum-wo-damp-c}
    }
    \subfigure[Experiment 1: Estimation of angular velocity over time.]{ %
    \centering
    \includegraphics[width=0.95\columnwidth]{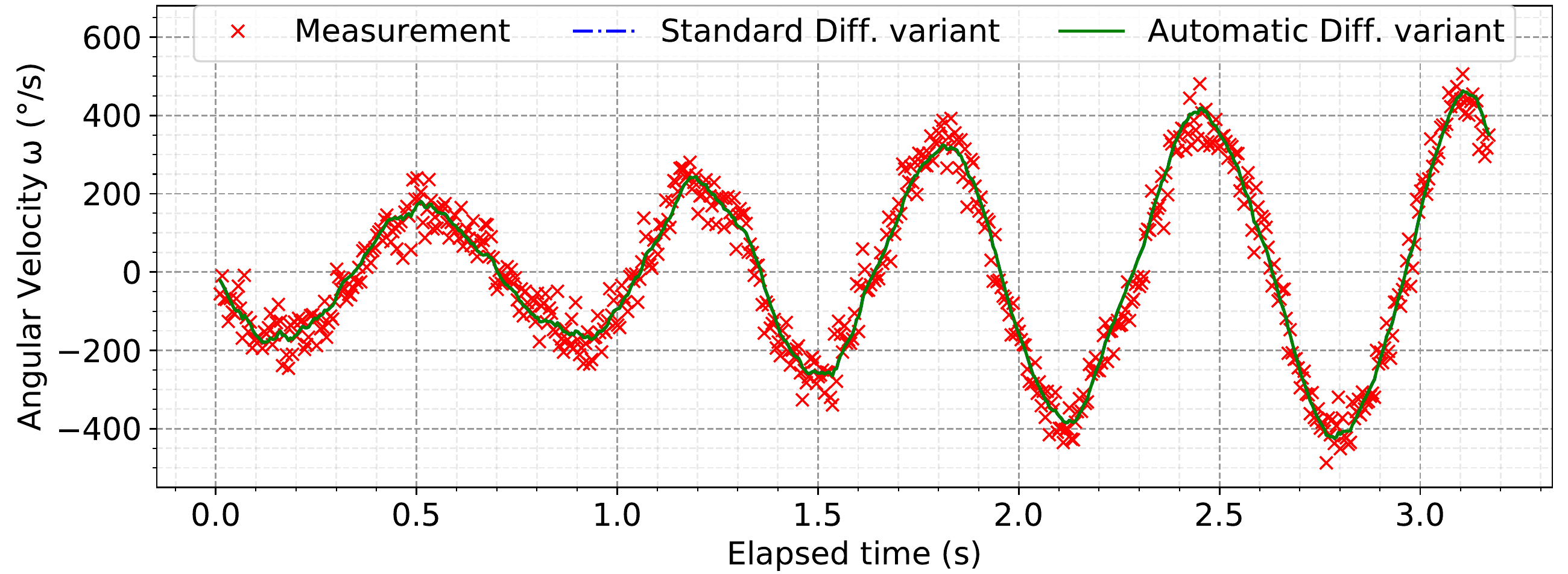}
    \label{fig:pendulum-wo-damp-b}
    }
    \quad\quad
	\subfigure[Experiment 2: Estimation of angular velocity over time.]{%
	\centering
    \includegraphics[width=0.95\columnwidth]{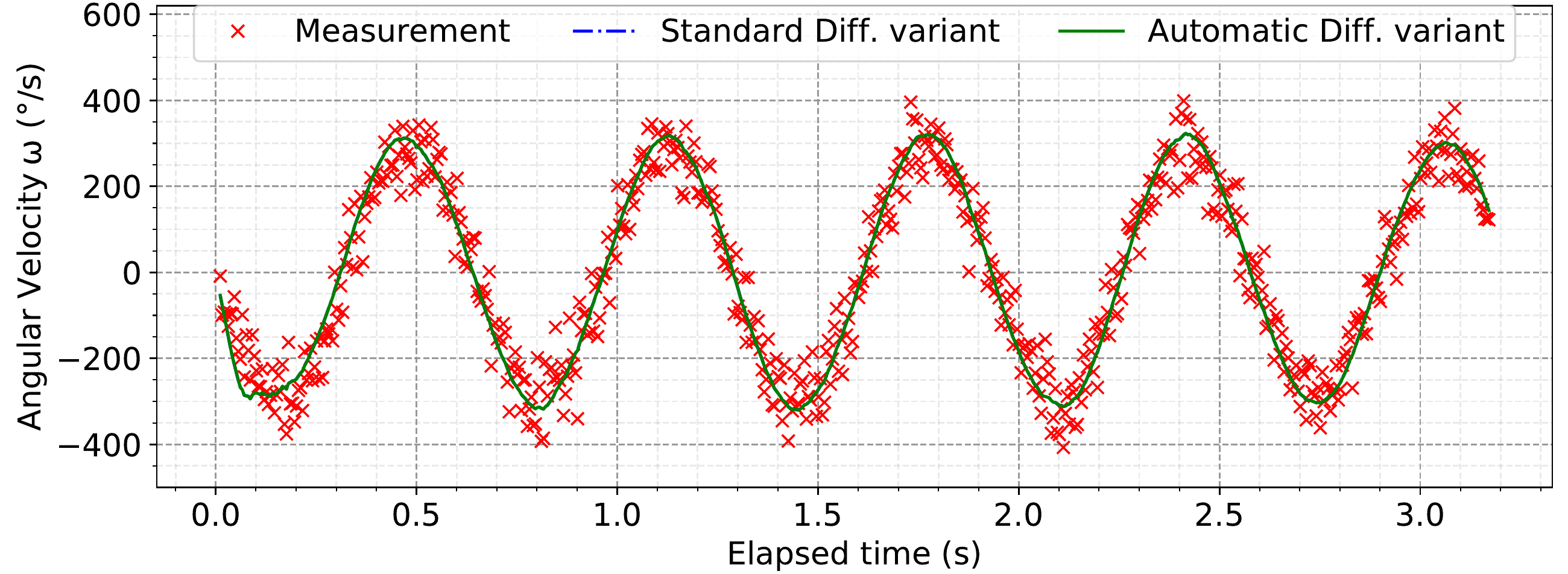}

    \label{fig:pendulum-wo-damp-d}
    }
    \caption{Evaluation of generated \kalman filter for estimation of the state a pendulum.
    	Sub-figures (a) and (c) correspond to an oscillation with initial displacement of $20^\circ$ 
    	and observable process noise of variance $0.005\,\mathrm{rad}^2/s^2$ in the angular velocity $\omega$ of the oscillation
    (notice the changes in maximum amplitude and frequency). The input to the filters is angular velocity from 
    a gyro with noise of variance $0.5\,\mathrm{rad}^2/s^2$. All generated filter variants manage to follow the noisy process
    with a mean square error of $0.0025\,\mathrm{rad}^2$ for the angle and $0.17\,\mathrm{rad}^2/s^2$ for the angular velocity.
    Sub-figures (b) and (d) correspond to a different pendulum simulation with initial displacement of $30^\circ$,
    measurement noise of variance $0.8\,\mathrm{rad}^2/s^2$. Examined filters were deliberately erroneously initialized
    with the knowledge of an initial displacement of $60^\circ$. Nevertheless, they manage to converge to accurate estimations
    about the state of the pendulum.}
    \label{fig:pendulum-wo-damp}
\end{figure*}

\subsection{Pendulum System}
\label{subsec:evaluation-pendulum}

In our first experiments we target the problem of state estimation of a 
physical system which simulates the oscillation of an ideal pendulum. 
The dynamics of the oscillation are presented in Equation~\ref{eq:PEND_1}.

We developed Newton descriptions similar to the one presented in Figure~\ref{fig:models-example}
to describe the pendulum with and without drag in order to use our implemented
Newton compiler back-end for the generation of \kalman filters.
The state vector of the system is $<\theta_t, \omega_t>$, composed of the oscillation
angle and the angular velocity, respectively.
The input of the filters is the angular velocity provided by a simulated gyroscope.
The measurements from the gyroscope are noisy and the noise was modelled, without
lack of generality, according to a Gaussian distribution. The angular rate of the
gyroscope is the only sensory input to the generated \kalman filters, which are
completely agnostic to the deviation of the estimated angle from the true one.
We generated two different variants of the filter, one that uses standard differentiation
and one that uses Automatic Differentiation.

We first evaluate a pendulum without drag, using its dynamics equation to 
simulate its oscillation in time and gather a trace of the state and sensor values.
We then use the generated \kalman filters to predict the time-evolution of the
pendulum state parameters in the presence of noisy input sensor measurements.
Figures~\ref{fig:pendulum-wo-damp-a} and~\ref{fig:pendulum-wo-damp-b} present 
an execution interval of the simulation, with initial pendulum displacement of 20$^{\circ}$ 
and observable process noise of variance $0.005\,\mathrm{rad}^2/s^2$ on the angular velocity $\omega$.
The sensor error distribution has zero mean value and variance equal to $0.5\,\mathrm{rad}^2/s^2$. 
The sensed values from the gyroscope are annotated in Figure~\ref{fig:pendulum-wo-damp-b} with red points,
combined with the pendulum angular velocity predicted by the two \kalman filter variants.
In Figure~\ref{fig:pendulum-wo-damp-b}, we provide the predicted pendulum angle by the two \kalman 
filters compared against the true value. We observe that both filter variants are capable of 
correctly estimating the oscillation with mean square error (MSE) of $0.003\,\mathrm{rad}^2$.

In our second experiment using the pendulum with no drag, we initialize the 
generated \kalman filters with a ``false'' initial angle displacement value. 
In this way, we evaluate their ability to estimate the state of the system 
when the initial state information are not accurate. 
The initial displacement was set to 30$^{\circ}$, while the 
filters were initialized with knowledge of 60$^{\circ}$ displacement. 
We provide the results of this experiment in Figure~\ref{fig:pendulum-wo-damp-c} and 
Figure~\ref{fig:pendulum-wo-damp-d}. We observe in Figure~\ref{fig:pendulum-wo-damp-c} 
that both filter variants are capable of converging to accurate predictions of
the system's state. For both filters, the MSE of the prediction of the oscillation 
angle $\theta$ was 0.0056$\,\mathrm{rad}^2$, while the MSE value for the prediction 
of the angular rate $\omega$ was 0.1757$\,\mathrm{rad}^2/s^2$.

\begin{figure}
	\subfigure[Estimation of oscillation angle over time.]{ %
    \includegraphics[width=0.95\linewidth]{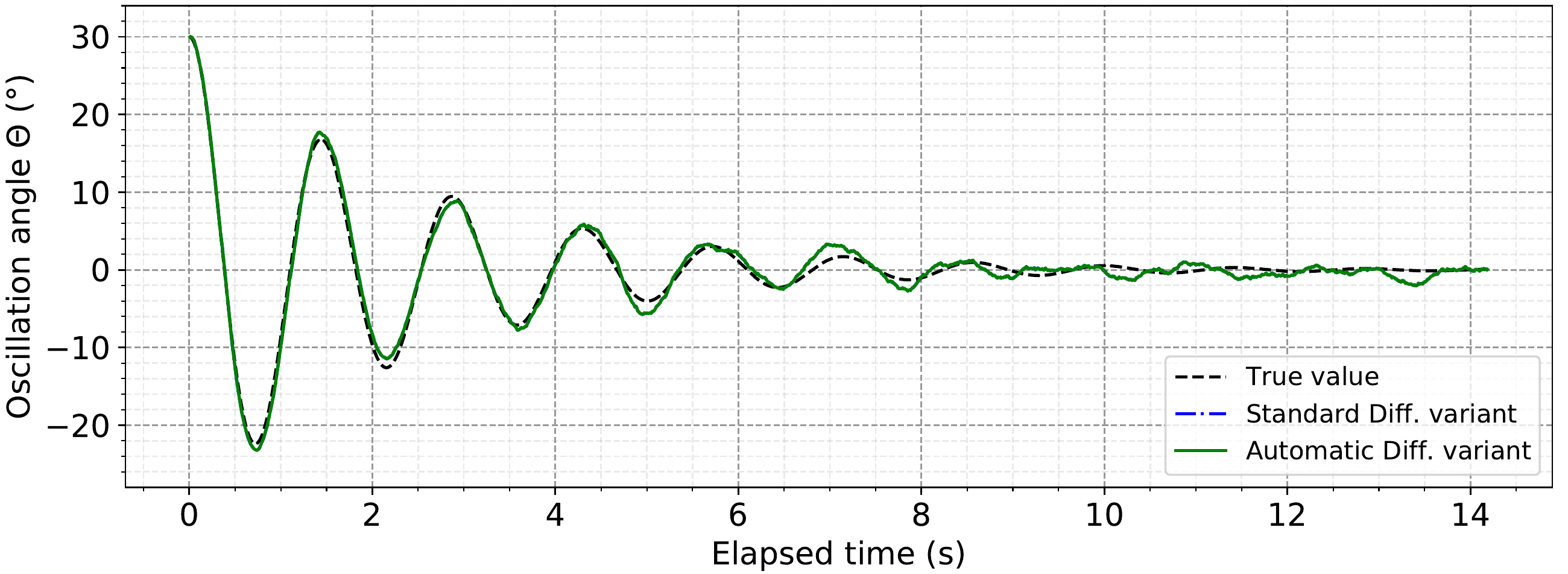}
    \label{fig:pendulum-with-damp-theta}
    }
    	\subfigure[Estimation of angular velocity over time.]{ %
    \includegraphics[width=0.95\linewidth]{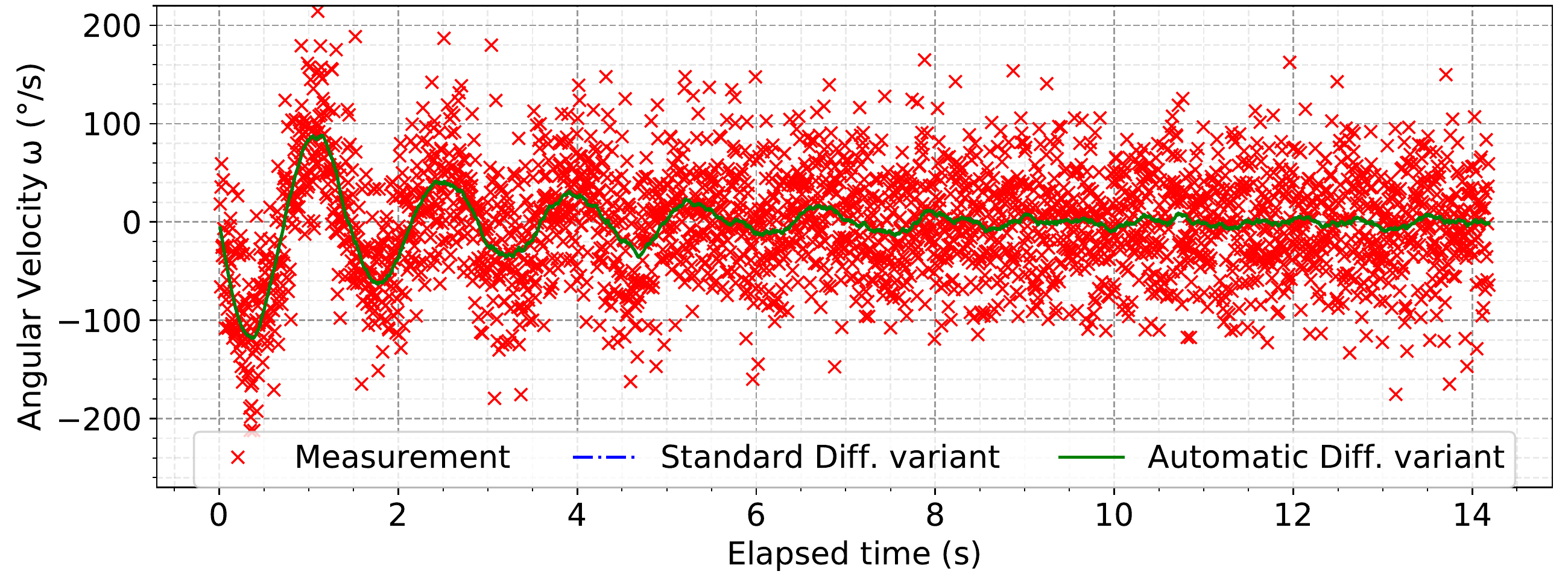}
    \label{fig:pendulum-with-damp-omega}
    }
    \caption{State estimation of a pendulum oscillation with drag. Initial displacement is 
    equal to 30$^\circ$ and measurement noise variance $0.8\,\mathrm{rad}^2/s^2$.
    Both generated filter variants, i.e., with standard or Automatic Differentiation, provide
    high accuracy estimates of the oscillation with diminishing amplitude.
    \label{fig:pendulum-with-damp}}
\end{figure}

We performed a third experiment for the state estimation of 
a pendulum with drag, length equal to 0.5\,$m$ and damping factor equal to 0.8\,$kg\,s^{-1}$.
The initial angle displacement was 30$^{\circ}$ and both filter variants
where provided correct information about this value. The results of this experiment
are shown in Figure~\ref{fig:pendulum-with-damp}. The measurement noise variance was
0.8$\,\mathrm{rad}^2/s^2$. 
We observe that both \kalman filter variants are capable of
making accurate predictions about the system state, despite the noise in measurements,
annotated with red points in Figure~\ref{fig:pendulum-with-damp-omega}. 
The MSE of the prediction for both filters was approximately 0.0002$\,\mathrm{rad}^2$ 
for the prediction of the oscillation angle $\theta$ and
0.0054$\,\mathrm{rad}^2/s^2$ for the prediction of the angular rate $\omega$.

\subsection{TurtleBot3 Robot}
To evaluate the effectiveness of the generated filters on the state estimation of a
complex cyber-physical system, we make use of the TurtleBot3 Burger~\cite{amsters2019turtlebot}, 
a robot built on open-source software. %
The TurtleBot3 Burger is a differential drive robot, i.e., it controls
its linear and angular velocities by actuating on a set of wheels with 
individual speed setpoints $v_{r}$ and $v_{l}$, for the right 
and the left wheel, respectively.
The general kinematic model of such a robot exhibits a non-holonomic constraint and
thus the process model, described by Dudek et al.~\cite{dudek2010computational}, 
is split in two special cases:

If $v_{r} = v_{l} = v$:
\begin{equation}
	\label{eq:tbotCaseA}
    \begin{bmatrix}
        x_{t+\delta t} \\ 
        y_{t+\delta t} \\ 
        \theta_{t+\delta t}
    \end{bmatrix}
    =
    \begin{bmatrix}
        x_t + v\ \cos(\theta_t) \delta t \\
        y_t + v\ \sin(\theta_t) \delta t \\ 
        \theta_t
    \end{bmatrix}
\end{equation}
and if $v_{r} = - v_{l} = v$:
\begin{equation}
	\label{eq:tbotCaseB}
    \begin{bmatrix}
        x_{t+\delta t} \\ 
        y_{t+\delta t} \\ 
        \theta_{t+\delta t}
    \end{bmatrix}
    =
    \begin{bmatrix}
        x_t \\
        y_t \\ 
        \theta_t + 2 v \delta t / l
    \end{bmatrix}
\end{equation}
where $x$, $y$ and $\theta$ are the robot's position and yaw in the fixed frame, 
$v$ is the velocity in the fixed frame, $\delta t$ is the time difference between
subsequent runs of the model and $l$ is the distance between the two wheels. The subscript
denotes a given point in time $t$.

We generated the robot process Newton invariant as a unified model
and hardcoded a C control flow command in the generated \textit{Predict} function 
to check the system's input, i.e. the velocities $v_{r}$, $v_{l}$, and perform the correct
operations according to either Equation~\ref{eq:tbotCaseA} or Equation~\ref{eq:tbotCaseB}. 
We instructed the measurement model to use the IMU and simulated localization information
from laser scanners of the robot. This information is 
treated as noisy information on the robot's position on the $XY$ plane, with
variance $0.1\,m^2$ for both directions. 
We simulated a stroll of the robot on the Gazebo~\cite{koenig2004design}
simulation platform and recorded the sensor measurements trace. 

\begin{figure}[]
\centering
    \subfigure[Actual path and path estimation of TurtleBot3.]{%
    \includegraphics[width=0.80\linewidth]{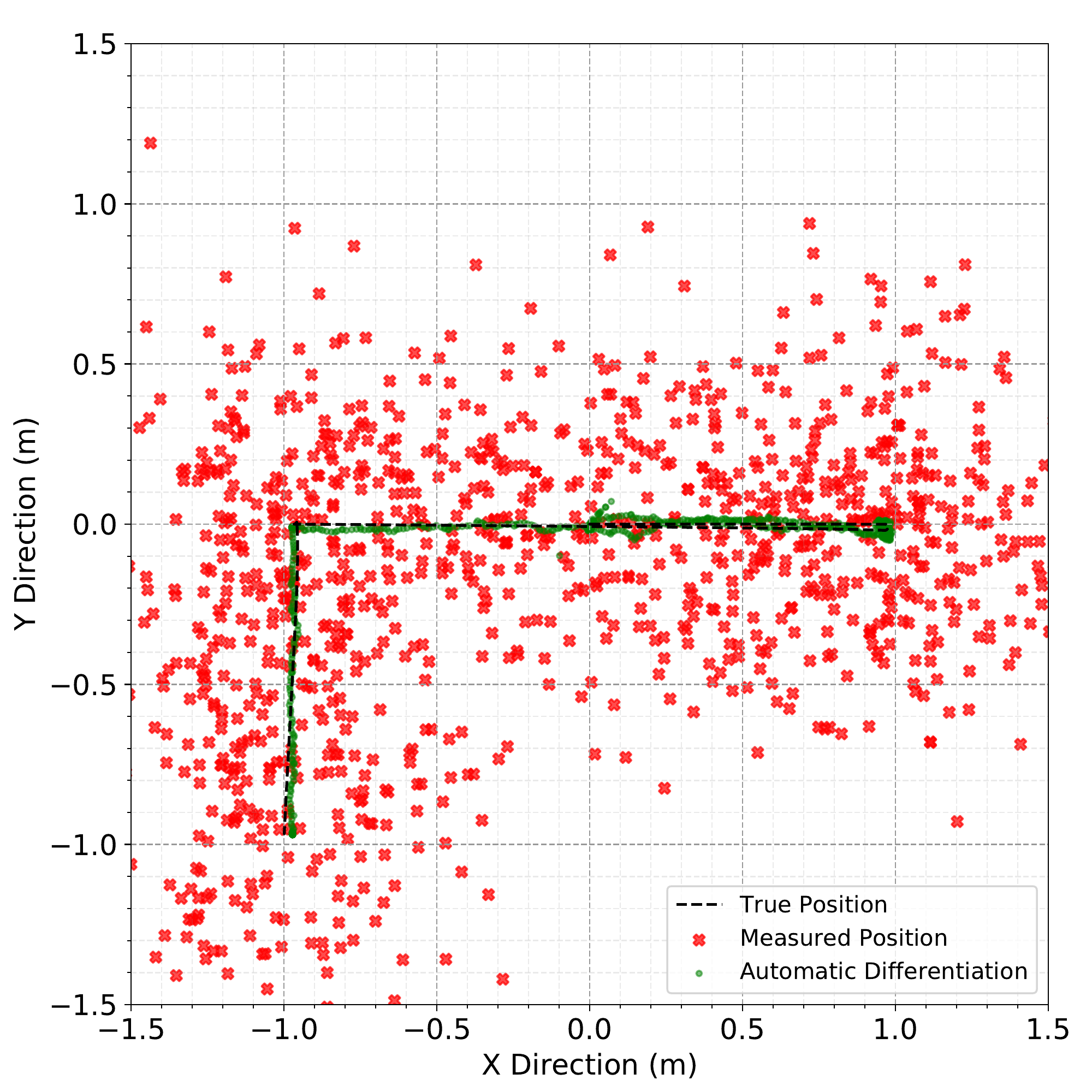}
    \label{fig:turtlebot3a}
    }
    \subfigure[Rotational movement estimation of TurtleBot3 using a gyroscope.]{
     \includegraphics[width=0.95\linewidth]{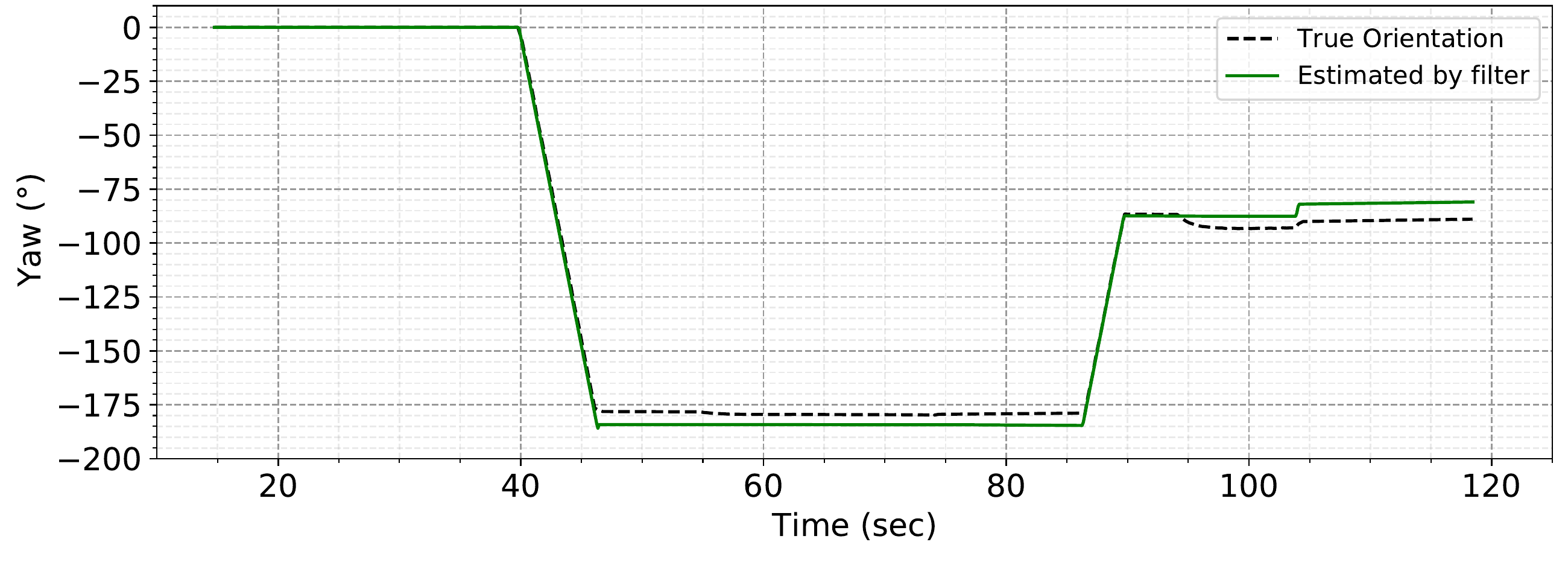}
    \label{fig:turtlebot3b}
    }
     \caption{Prediction of the path of the TurtleBot3 Burger robot. It starts at position $(0,0)$ 
     facing the positive values of x axis. It moves straight to $(1,0)$, turns 180$^\circ$ left, 
     continues to $(-1,0)$, turns 90$^\circ$ left and finally moves to $(-1,-1)$ as shown in Sub-figure (a).
     Sub-figure (b) shows the actual and estimated value of the yaw of the robot using a generated
     \kalman filter.}
     \label{fig:turtlebot3}
\end{figure}

We utilized the recorded trace to evaluate the effectiveness of our generated
\kalman filter with Automatic Differentiation in predicting the state vector 
$<x_t, y_t, \theta_t>$ of the robot. Figure~\ref{fig:turtlebot3} presents 
the estimation results for the position (Figure~\ref{fig:turtlebot3a})
and yaw (Figure~\ref{fig:turtlebot3b}) of the robot, both represented using green lines.
The recorded measurements from the sensor of the robot are annotated in 
Figure~\ref{fig:turtlebot3a} with red points.
The robot started its stroll at position $(0,0)$ facing towards the positive side
of the x axis. We instructed it to to move straight to position $(1,0)$, 
turn 180$^\circ$ right, continue to $(-1,0)$, 
turn 90$^\circ$ left and finish its movement at position $(-1,-1)$.
The dashed black line in Figure~\ref{fig:turtlebot3a} corresponds to the 
actual trajectory of the robot and we can observe the high estimation
accuracy of our generated \kalman filter, despite the noise in the
measured values. The average Euclidean error of the translational movement 
of the robot is 0.0185$\,m$ whereas the average error of the 
rotational movement (Figure~\ref{fig:turtlebot3b}) is 4.72$^\circ$.

\subsection{Code Size \& Performance Evaluation}
Additionally to evaluating the accuracy of the
generated filters, it is also important to 
quantify their memory and computational requirements.
Thus, we profiled the generated filters for state estimation of a pendulum without drag, 
as presented in Section~\ref{subsec:evaluation-pendulum}.
We choose RISC-V as the target reference 32-bit RISC CPU architecture and examine
different extensions of the instruction set architecture (ISA) to take into account 
embedded systems of different computational competency.
Starting from the RV32I base integer ISA, which provides only integer addition/logical 
operations, we also examine RV32IM extension with integer multiplication and division 
instructions and RV32IMF/RV32IMFD extensions for single/double precision floating-point 
arithmetic operations.

The generated \kalman filter source code contains double precision variables and has been compiled
using GCC 8.2.0. %
We examined both variants of standard and Automatic Differentiation and 
benchmarked the generated filters for an increasing amount of input workload 
by varying the amount of total sensor inputs that
are being processed for estimation of the state.
For this experiment, we examine inputs of 10, 100 and 500 sensor measurements of
the angular rate $\omega$ of the pendulum.

Both filters were compiled with the `-O3' optimization option and we evaluated the binary size 
of the generated \kalman filters using the \textit{riscv32-elf-size} tool.
We additionally evaluated the dynamic instruction count for the execution of the \kalman filters
using Sunflower~\cite{stanley2007sunflower}, an open-source embedded system micro-architectural 
emulator. %
The input sensor samples have been hardcoded in the source in order to decouple our measurements
from the required instructions for disk I/O.

\begin{figure}[]
	\subfigure[Binary text size in kB.]{
    \includegraphics[width=0.42\columnwidth]{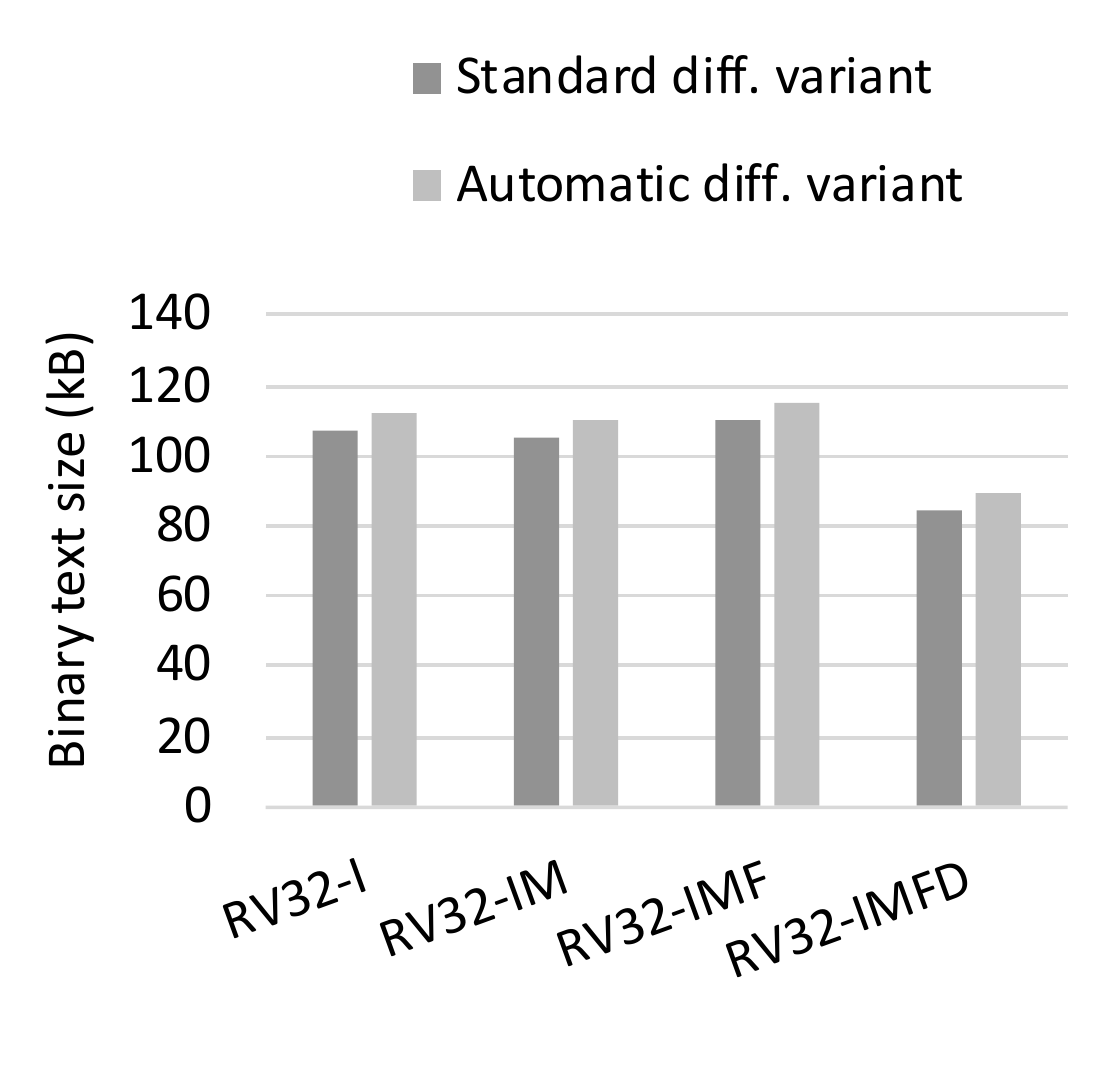}
    \label{fig:text-size-sd-vs-ad-O3}
    }
    \subfigure[Dyn. instruction count (log scale).]{
    \includegraphics[width=0.52\columnwidth]{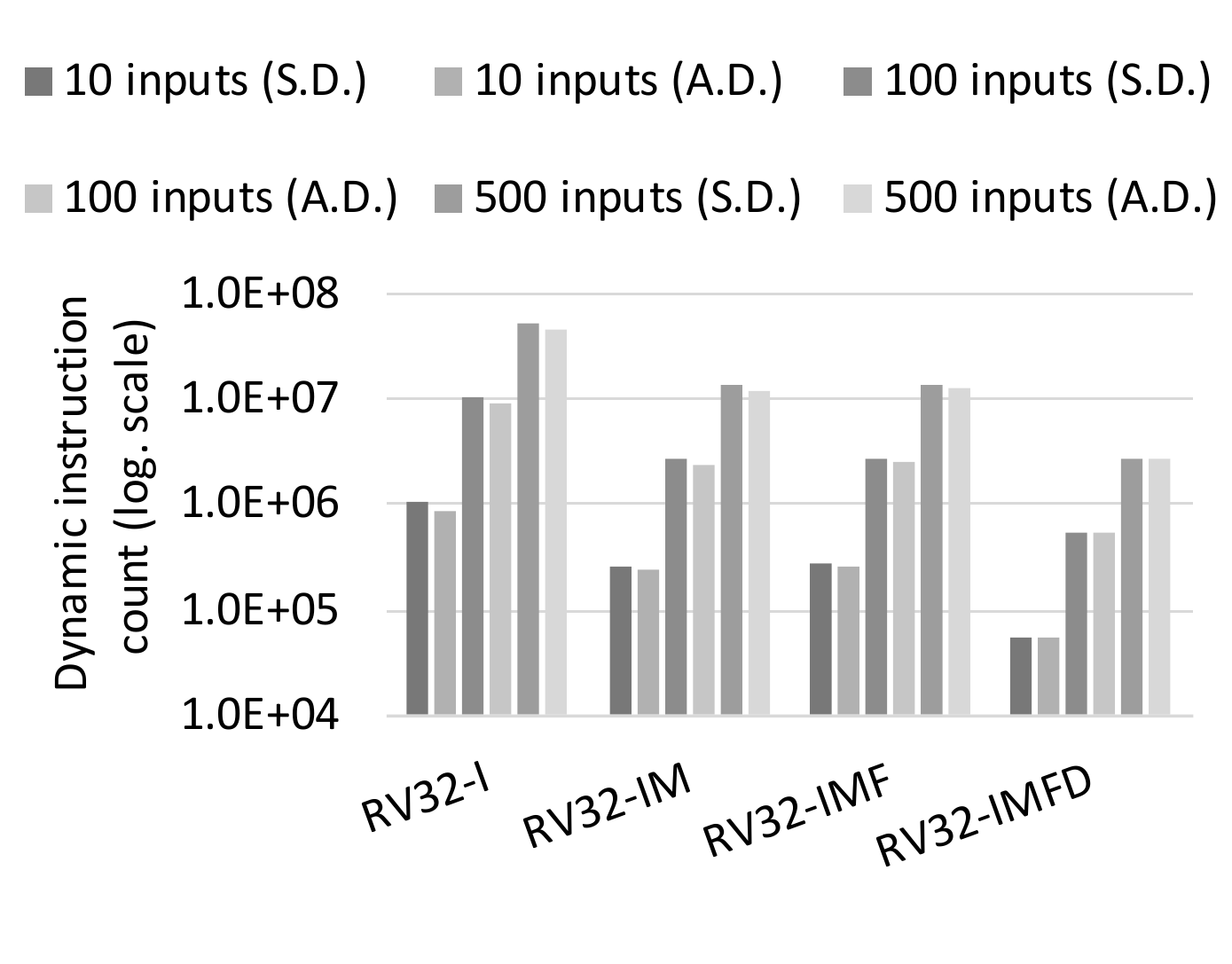}
    \label{fig:ni-sd-vs-ad-O3}
    }
    \caption{Profiling of the compiled binaries of the generated filters' source code
    for various RISC-V ISA extensions. Sub-figure (a) shows the size of the `text'
    segment of the binaries, showing a slight increase for filters with Automatic Differentiation.
    Sub-figure (b) shows the required instructions for the processing of input sensor vectors, 
    where the filters with Automatic Differentiation achieve
    significant reduction compared to standard differentiation.
    }
    \label{fig:sf-profiling}
\end{figure}

The size of the `text' segment of the filter binaries, 
for the different ISA extensions is illustrated in Figure~\ref{fig:text-size-sd-vs-ad-O3}. 
The dynamic instructions count for the completion of the filters' execution for varying input size 
is presented in Figure~\ref{fig:ni-sd-vs-ad-O3}. 
Standard and Automatic Differentiation filter variants are annotated using `S.D.' and `A.D', respectively.
Each bar of Figure~\ref{fig:ni-sd-vs-ad-O3} corresponds to a different workload, i.e. different total
number of input sensor vectors processed by the filter.

We observe that in all cases the filter variant with Automatic Differentiation is slightly larger
in size but requires less instructions for its completion (note that the Y-axis scale of
Figure~\ref{fig:ni-sd-vs-ad-O3} is logarithmic). 
Automatic differentiation results in an average gain of more than 7.5\% in the required
instructions for the filter execution, exceeding 16\% at its maximum in the case of RV32I ISA.
The tradeoff is an average increase of 4.7\% in the `text' segment of the compiled binary.

We attribute the increase in `text' size to (i) the expansions of the \kalman filter 
equations to their SSA form and (ii) the inclusion of the SSA form of the Reverse 
Mode AutoDiff (Section~\ref{section:autodiff_ekf}). 
The reduction of the dynamic instructions is attributed 
to the usage of Automatic Differentiation, which calculates the partial 
derivatives in the rows of the Jacobian matrices in a single function evaluation 
(Section~\ref{sec:synthesized_api}). Conversely, standard differentiation 
needs to evaluate the required functions two times more for each partial 
derivative of a row.

\section{Related Research}
Automating \kalman filter design and implementation process is a
task of high importance for engineers in various domains.  As a
result, MATLAB and GNU Octave both provide libraries that assist
the design of \kalman filters~\cite{grewal2014kalman, sarkka2013bayesian}
and MATLAB provides the ability for automated generation of C/C++
source code for filters.  A python package for \kalman filtering
is also available~\cite{labbe2015kalman}, offering a range of filters
and great documentation, but focusing on pedagogy rather than
implementation efficiency.  These solutions are either proprietary
or are not directly applicable to the majority of embedded applications
as these require small code size and small memory usage.

The most prominent related research is the
AutoFilter~\cite{whittle2004automating, richardson2006flexible}, a
closed-source tool which supports code generation for the LKF, the
linearized filter, and the EKF.  AutoFilter offers abstractions for
transformations such as linearization and transforming the system
from continuous to discrete.  The generated code can be in the form
of C/C++, Modula-II source code, or a MATLAB script.  AutoFilter
relies on an input grammar, which supports the definition of
constants, input data, datatypes, vectors, matrices and
distributions~\cite{whittle2004automating}.  The target model in
this grammar is declared as a list of equations.  This approach
favors the description of non-linear models but is unintuitive in
the case of linear systems.  Furthermore, although the authors claim
that the tool supports differential equations, the input grammar
merely equates each differential of the state variables with an
algebraic expression.

AutoFilter and the automated solutions and libraries described
above, all assume a static time difference between execution steps.
In real-world embedded systems however, the time difference between
consecutive readings from one sensor may vary and readings across
multiple sensors are often at different timestamps. In contrast to
all prior work, the method we present in this work makes no such
assumptions about time steps. Our method exploits information about
the physics of a system, is generalizable beyond \kalman filters,
and builds on recent results in Automatic
Differentiation~\cite{baydin2017automatic, margossian2019review}
to automate the generation of the Jacobians in the non-linear EKF.

\section{Conclusion and Future Work}
This article presents an advance in the state of the art
in automated synthesis of state estimation and sensor fusion
algorithms.  The method we present starts from a specification of
the physics of an embedded sensor-driven system and its environment.
It generates, as output, C code with small code and memory
footprint, suitable for deployment of ultra-low-power microcontrollers.
The automation of the method by its implementation within a compiler
for a physics specification language reduces the time-consuming
and potentially error-prone process of designing and updating state
estimation algorithms such as linear and extended \kalman filters.

The method we present is easily extended to support other state
estimation algorithms, such as the unscented \kalman filter and the
particle filter, or different approaches to filter subcomponents,
such as using an information matrix instead of a covariance matrix
\cite{terejanu2013discrete}, enabling rapid comparison between
multiple filter variants for the same embedded solution.  Additionally,
the implementation of the method within a compiler for a physical
system specification language makes it possible to add more static
compile-time analysis for the output system, optimizing multiple
parts of the resulting code in terms of efficiency, complexity and
code size as well as providing this information to the user before
deployment.

Our implementation exploits recent advances in Reverse-Mode Automatic
Differentiation of program sequences, borrowed from the world of
machine learning, to automate the generation of the partial derivatives
of the state equation for the Jacobian matrices for the extended
\kalman filter. Using descriptions of physical systems of a range
of complexities, we evaluate and validate the generated filters in
terms of accuracy, stability, convergence, and run-time requirements
for deployment in resource-constrained environments.

\bibliographystyle{IEEEtran}
\bibliography{references}
\end{document}